\documentclass{article}
\usepackage[utf8]{inputenc}
\usepackage{epsfig,latexsym,amsfonts,amsmath,amsthm,amssymb,amsbsy,multirow,slashed,wasysym,textcomp,comment,mathtools,cancel,bbold,cite,datetime}
\usepackage{geometry}
 \usepackage{pgfplots}
 \usepgfplotslibrary{patchplots}
 \pgfplotsset{compat=newest}

\theoremstyle{definition}

\DeclareGraphicsExtensions{.pdf,.png}
\usetikzlibrary{patterns}
\usepackage{tikz}
\usetikzlibrary{arrows.meta}
\tikzstyle{bag} = [align=center]
\usetikzlibrary{decorations.pathmorphing}
\usepackage[normalem]{ulem}

\begin{document}
 \begin{titlepage}
 \begin{flushright}{${}$}
 \end{flushright}
	 \vspace{1.5cm}
\begin{center}
  \baselineskip=13pt {\Large \scshape{HPS meets AMPS:\\[4mm] How soft hair dissolves the firewall}}
   \vskip1.5cm 
   {\large Sabrina Pasterski}${}^\diamondsuit$
   {\large and Herman Verlinde}${}^\blacklozenge$\\[8mm]
   {\em${}^\blacklozenge$ Physics Department, Joseph Henry Laboratories}\\[3mm]
\noindent{${}^\diamondsuit$ \it Princeton Center for Theoretical Science}\\[3mm]
{\em Princeton University, Princeton, NJ 08544, USA}

\end{center}

\vspace{2cm}

\begin{abstract}
\addtolength{\baselineskip}{.5mm}
We build on the observation by Hawking, Perry and Strominger that a global black hole space-time supports a large number of soft hair degrees of freedom to shed new light on the firewall argument by Almheiri, Marolf, Polchinski, and Sully. We propose that the soft hair Goldstone mode is encoded in a classical transition function that connects the asymptotic and near horizon region. The entropy carried by the soft hair is part of the black hole entropy and encoded in the outside geometry. We argue that the infalling observer automatically measures the classical value of the soft mode before reaching the horizon and that this measurement implements a code subspace projection that enables the reconstruction of interior operators.  We use the soft hair dynamics to introduce an observer dependent notion of the firewall and show that for an infalling observer it recedes inwards into the black hole interior: the observer never encounters a firewall before reaching the singularity.  Our results indicate that the HPS black hole soft hair plays an essential role in dissolving the AMPS firewall.

\end{abstract}

\vfill

\date{December 2020}

\end{titlepage}

\clearpage 

\addtolength\baselineskip{1mm}
\tableofcontents

\addtolength\parskip{.7mm}
\addtolength\baselineskip{-.5mm}
\def\is{\! & \! = \! & \!}
\def\half{{1\over 2}}
\numberwithin{equation}{section}
\def\sub{\scriptscriptstyle}
\def\ip{${\mathcal I}^+$}
\def\calr{{\cal R}}
\def\e{{\epsilon}}
\def\g{{\gamma}}
\def\cs{{\cal S}}
\def\S{\Sigma }
\def\s{\sigma }
\def\sz{\sigma^{0} }
\def\Psz{\Psi^{0} }
 \def\p{\partial}
 \def\bz{{\bar z}}
 \def\cT{8\pi G_N {\mathcal T}}
\def\0{{(0)}}
\def\1{{(1)}}
\def\2{{(2)}}
 \def\cL{{\cal L}}
\def\co{{\cal O}}\def\cv{{\cal V}}
\def\n{\nabla}
\def\ci{{\mathcal I}}
\def\ipp{${\mathcal I}^+_+$}
\def\<{\langle }
\def\>{\rangle }
\def\[{\left[}
\def\]{\right]}
\def\bw{{\bar w}}
\def\h{{h^i_i}}
\def\t{{\rm{trace}}}
\def\o{\omega }
\def\ra{\bigr\rangle}
\newcommand{\non}{\nonumber}
\renewcommand{\O}{\Omega}
\renewcommand{\L}{\Lambda}
\newcommand{\bigO}{\mathcal{O}}
\newcommand{\sech}{\mbox{sech}}
\newcommand{\tn}{\tilde{n}}
\newcommand{\W}{\mathcal{W}}
\newcommand{\tr}{\mbox{tr}}
\newcommand{\Del}{\nabla}
\newcommand{\hs}[1]{\mbox{hs$[#1]$}}
\newcommand{\w}[1]{\mbox{$\W_\infty[#1]$}}
\newcommand{\bif}[2]{\small\left(\!\!\begin{array}{c}#1 \\#2\end{array}\!\!\right)}
\renewcommand{\u}{\mathfrak{u}}
\newcommand{\scriplus}{\mathcal{I}^+}
\renewcommand{\epsilon}{\varepsilon}
\def\dim#1{\lbrack\!\lbrack #1 \rbrack\!\rbrack }
\newcommand{\chichi}{\chi\!\cdot\!\chi}
\newcommand\snote[1]{\textcolor{magenta}{[S:\,#1]}}

\renewcommand{\theequation}{\thesection.\arabic{equation}}
   \makeatletter
  \let\over=\@@over \let\overwithdelims=\@@overwithdelims
  \let\atop=\@@atop \let\atopwithdelims=\@@atopwithdelims
  \let\above=\@@above \let\abovewithdelims=\@@abovewithdelims
\renewcommand\section{\@startsection {section}{1}{\z@}%
                                   {-3.5ex \@plus -1ex \@minus -.2ex}
                                   {2.3ex \@plus.2ex}%
                                   {\normalfont\large\bfseries}}

\renewcommand\subsection{\@startsection{subsection}{2}{\z@}%
                                     {-3.25ex\@plus -1ex \@minus -.2ex}%
                                     {1.5ex \@plus .2ex}%
                                     {\normalfont\bfseries}}

\newcommand{\Tr}{\mbox{Tr}}
\renewcommand{\H}{\mathcal{H}}
\newcommand{\SU}{\mbox{SU}}
\newcommand{\chiu}{\chi^{{\rm U}(\infty)}}
\newcommand{\ff}{\rm f}
\linespread{1.3}

\def\bfR{{\mbox{\textbf R}}}
\def\gzz{\gamma_{z\bz}}
\def\vx{{\vec x}}
\def\p{\partial}
\def\po{$\cal P_O$}
\def\cN{{\cal H}^+ }
\def\N{${\cal H}^+  ~~$}
\def\G{\Gamma}
\def\l{{\ell}}
\def\ch{{\cal H}^+}
\def\Q{{\hat Q}}
\def\T{\hat T}
\def\C{\hat C}
\def\zet{z}
\def\scc{\mbox{\small $\hat{C}$}}
\def\Aa{{\mbox{\scriptsize \smpc \sc a}}}
\def\aalpha{{\mbox{\scriptsize {\smpc  $\alpha$}}}}
\def\cC{{\mbox{\scriptsize {\smpc \sc c}}}}
\def\cCb{{\mbox{\scriptsize \smpc \sc c'}}}
\def\sS{{\mbox{\scriptsize {\smpc \sc s}}}}
\def\Bb{{\mbox{\scriptsize \smpc \sc b}}}
\def\Hh{{\mbox{\scriptsize \smpc \sc h}}}
\def\oO{{}} 
\def\bfC{\mbox{{\textbf C}}}
\def\nonu{\nonumber}
\def\im{{\rm i}}
\def\tr{{\rm tr}}
\def\be{\bea}
\def\ee{\eea}

\def\spc{\hspace{.5pt}}

\def\bea{\begin{eqnarray}}
\def\eea{\end{eqnarray}}
\def\is{\! & \! = \! & \!}
\def\half{{\textstyle{\frac 12}}}
\def\cL{{\cal L}}
\def\halfi{{\textstyle{\frac i 2}}}

\def\ba{\begin{eqnarray}}
\def\ea{\end{eqnarray}}

\def\delbar{\overline{\partial}}
\newcommand{\smpc}{\hspace{.5pt}}
\def\nspc{\!\spc\smpc}
\def\uU{\mbox{\textit{\textbf{U}}\spc}}
\def\uV{\mbox{\textit{\textbf{U}\spc}}}
\def\tT{\mbox{\textit{\textbf{T}}\spc}}
\def\bfC{\mbox{\textit{\textbf{C}}}}
\def\bn{\mbox{\textit{\textbf{n}\!\,}}}
\def\pP{\mbox{\textit{\textbf{P}\!\,}}}
\def\rR{{\textit{\textbf{R}\!\,}}}

\def\im{{\rm i}}
\def\tr{{\rm tr}}

\def\ra{\bigr\rangle}
\def\la{\bigl\langle}
\def\li{\bigl |\spc}
\def\ri{\bigr |\spc}

\def\nonu{\nonumber}

\enlargethispage{\baselineskip}

\setcounter{tocdepth}{2}
\newpage
\addtolength{\baselineskip}{.3mm}
\addtolength{\parskip}{.3mm}
\addtolength{\abovedisplayskip}{.9mm}
\addtolength{\belowdisplayskip}{.9mm}
\renewcommand\Large{\fontsize{15.5}{16}\selectfont}

\newcommand{\newsubsection}[1]{
\vspace{.6cm}
\pagebreak[3]
\addtocounter{subsubsection}{1}
\addcontentsline{toc}{subsection}{\protect
\numberline{\arabic{section}.\arabic{subsection}.\arabic{subsubsection}}{#1}}
\noindent{\arabic{subsubsection}. \bf #1}
\nopagebreak
\vspace{1mm}
\nopagebreak}
\renewcommand{\footnotesize}{\small}
\section{Introduction}
\vspace{-2mm}

The question of how black holes process quantum information remains one of the central unresolved problems in theoretical physics. A famous manifestation of this puzzle is the AMPS firewall paradox~\cite{Almheiri:2012rt}. Unitary quantum evolution and the existence of a smooth horizon are known to be mutually consistent for pre-Page time black holes with von Neumann entropy below the Bekenstein-Hawking bound. For maximally mixed black hole states, however, one encounters a conundrum: how can an evaporating post-Page time black hole decrease its von Neumann entropy, while the Hawking process and smoothness of the horizon require that it keeps generating new entanglement with its immediate surroundings?

The AMPS argument makes a number of implicit assumptions. Notably, it assumes that a semi-classical black hole space-time, as seen by an infalling observer, can be in a maximally mixed state that saturates the BH bound. This premise receives support from the no hair theorem, which states that the only information accessible from outside the black hole are its mass, charge and angular momentum. The no hair theorem is a true statement about the local black hole geometry. However, it does not hold as a global statement: a global black hole space-time carries a large amount of so-called soft-hair degrees of freedom associated with the asymptotic BMS symmetry group. As first argued by Hawking, Perry and Strominger, this soft hair must be viewed as part of the quantum information carried by the black hole~\cite{Hawking:2015qqa,Hawking:2016msc,Hawking:2016sgy,Strominger:2017aeh}. 

Follow-up work on the HPS proposal mostly centered on its potential implications  on the black hole information paradox and on the question of whether the soft hair could be responsible for the BH entropy \cite{Mirbabayi:2016axw,Bousso:2017dny,Haco:2018ske,Donnay:2018ckb,Haco:2019ggi}.
Thus far, however, little attention has been paid to the implications of soft hair on the firewall paradox. The purpose of this work is to fill this gap.\footnote{The relevance of soft modes to the firewall was also pointed out recently in \cite{Nomura:2018kia,Nomura:2019dlz}} Can we use the arguments of HPS to get around the reasoning of AMPS? Monogamy of entanglement has different implications if there are soft modes that both the black hole and Hawking radiation are necessarily entangled with. As we will argue, the presence of black hole soft hair is indeed sufficient to avoid the conclusion that infalling observers must experience a firewall.

\subsection{Argumentation} 
\vspace{-2mm}

We start with an outline of our reasoning for why the HPS proposal invalidates the AMPS conclusion.

\vspace{-2.5mm}

\newsubsection{~Firewall argument and no hair theorem}

The firewall argument~\cite{Almheiri:2012rt,Marolf:2013dba} and the recent Island prescription~\cite{Penington:2019npb,Almheiri:2019yqk,Almheiri:2019psf,Almheiri:2019hni} rely on the physical assumption that the black hole naturally evolves into a maximally mixed state, compatible with its macroscopic quantum numbers.  This premise is a special case of the general physical fact that any non-isolated quantum system interacting with a large environment ultimately evolves into a thermal mixed state specified by the macroscopic characteristics of the system. What is special about black holes, however, is that they appear to have so few macroscopically distinguishable classical properties relative to their total number of micro-states. 

The number of black hole states with the same macroscopic quantum numbers is given by the Bekenstein-Hawking entropy. Once the von Neumann entropy of the black hole saturates the BH bound, it can no longer grow more entangled with its environment. This leads to an apparent obstruction to continuing the Hawking process in a manner that preserves the smoothness of the black hole horizon. This is the firewall argument. The Island proposal aims to evade its conclusion via the postulate that the interior region of the black hole is contained inside the entanglement wedge of the Hawking radiation \cite{Penington:2019npb,Almheiri:2019yqk,Almheiri:2019psf,Almheiri:2019hni}. 

Both arguments rely on a coarse grained notion of a black hole that ignores the fine-grained properties of the global black hole space-time. The original no-hair theorem \cite{Israel:1967wq,Israel:1967za} only implies uniqueness up to local diffeomorphisms. It is now understood that global black hole space-times support a large family of soft Goldstone modes in the form of non-trivial diffeomorphisms that relate the near horizon and the asymptotic regions. These diffeomorphism modes are associated to the presence of an asymptotic symmetry group (ASG) of the black hole space-time. In asymptotically flat 3+1D black hole space-times, the ASG is the BMS group~\cite{Bondi:1962px,Sachs:1962wk}; in 2+1D AdS black holes they generate the Virasoro symmetry group of the dual CFT~\cite{brown1986}; in 1+1D near-AdS black hole models, the ASG simplifies to reparametrizations of the asymptotic time-coordinate~\cite{Almheiri:2014cka,Maldacena:2016upp,Jensen:2016pah,Engelsoy:2016xyb}. The presence of the asymptotic symmetry group implies that black holes in fact do carry soft hair degrees of freedom in the form of Goldstone modes associated with the breaking of the asymptotic symmetries due to the presence of the black hole horizon. 

\newsubsection{~Physical properties of soft hair}

The soft hair degrees of freedom can carry a sizable amount of entropy. Since this entropy only exists by virtue of the presence of the black hole, it should be counted as part of the black hole entropy. Previous studies have attempted to show that the soft hair is sufficiently abundant to account for all the BH entropy~\cite{Haco:2018ske,Donnay:2018ckb,Haco:2019ggi}. For our context,  however, it will not be essential whether this holds or not. For our arguments, it is sufficient to assume that the amount of entropy encoded in the soft hair is large compared to the entropy of the low energy QFT degrees of freedom contained inside of the black hole.

The key assertions that make our story work, are that the soft hair black hole degrees of freedom: 

\vspace{-2mm}

\begin{enumerate}
\addtolength\parskip{-1.75mm}
\item{ are a classical and measurable property of the global black hole space-time}
\item{are invisible to an asymptotic observer or to a local observer at the horizon}
\item{can be measured by an observer falling in from infinity into the black hole}
\item{are an exponentially sensitive probe of infalling matter into the black hole}
\item{carry a large amount of entropy compared to the entropy of the interior low energy QFT}
\item{are projected onto a low entropy state after repeated measurements by an infalling observer }
\item{accumulate more and more entropy during the Hawking evaporation process.}
\addtolength\parskip{1.75mm}
\end{enumerate}

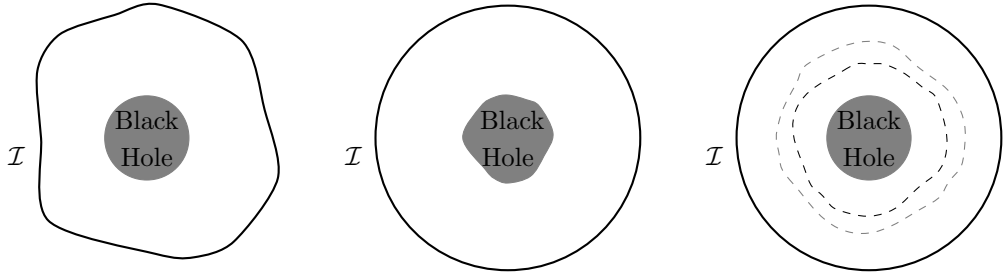
\begin{figure}[hbtp]
\centering
\begin{tikzpicture}[scale=.8,smooth cycle] 
  \draw[xshift=6cm,scale = 2.2,thick,black] plot[tension=1]
    coordinates{(-1,0)(0,1)(1,0)(0,-1)};   
    \draw[xshift =3.8cm] node[below left]{${\cal I\,}$};
  \draw[xshift=6cm,fill, gray,scale=.75] plot[tension=0.6] 
    coordinates{(-1,0) (-.6,.7)(-.3,.9)(.0,.95) (.4,0.85) (.7,.7)(1,0.1)(.8,-.4) (.5,-0.85) (0,-1)(-.3,-.9)(-.5,-.7)};
       \draw[xshift=6cm] node[above] {\ Black};
  \draw[xshift=6cm] node[below] {Hole}; 
  \draw[xshift=0cm,thick,black,scale=2.25,rotate=40] plot[tension=0.5] 
    coordinates{(-.95,0) (-.6,.5)(-.3,.9)(.0,1) (.4,0.85) (.7,.7)(1,0.1)(.8,-.4) (.55,-0.85) (0,-1)(-.3,-.9)(-.5,-.7)};
    \draw[xshift =-1.8cm] node[below left]{${\cal I\,}$};
  \draw[fill, gray,xshift=0cm,scale = .7] plot[tension=1] 
    coordinates{(-1,0) (0,1)(1,0)(0,-1)};   
  \draw node[above] {Black}   ;
  \draw node[below] {Hole}  ; 
   \draw[xshift=12cm,thick,black,scale = 2.2] plot[tension=1] 
    coordinates{(-1,0) (0,1)(1,0)(0,-1)}; 
  \draw[xshift=12cm,dashed,gray,scale=1.6,rotate=80] plot[tension=0.8]  coordinates{(-1,0) (-.6,.7)(-.3,.9)(.0,.95) (.4,0.85) (.7,.7)(1,0.1)(.8,-.4) (.5,-0.85) (0,-1)(-.3,-.9)(-.6,-.7)}; 
  \draw[xshift=12cm,dashed,black,scale=1.3,rotate=-20] plot[tension=0.8]  coordinates{(-1,0) (-.6,.7)(-.3,.9)(.0,.95) (.4,0.85) (.7,.7)(1,0.1)(.8,-.4) (.5,-0.85) (0,-1)(-.3,-.9)(-.6,-.7)}; 
    \draw[xshift =9.8cm] node[below left]{${\cal I\,}$};
\draw[fill, gray,xshift=12cm,scale = .7] plot[tension=1] 
    coordinates{(-1,0) (0,1)(1,0)(0,-1)};
       \draw[xshift=12cm] node[above] {Black};
  \draw[xshift=12cm] node[below] {Hole}; 
\end{tikzpicture}
\caption{\addtolength{\baselineskip}{-.5mm}Schematic depiction of three ways of storing the soft hair degrees of freedom. The no hair coordinate system that uniformizes the black hole geometry is incompatible with the coordinate system that uniformizes the asymptotic geometry ${\cal I}$.  If one tries to describe the global space-time with a single coordinate patch, either the asymptotic geometry (left) or the black hole geometry (middle) is not uniformized and carries the soft hair.  By using two coordinate patches, one can uniformize both the black hole and the asymptotic region. The soft hair modes are encoded in the transition function between the two regions.}
\label{fig:softhair}
\end{figure}

\vspace{-2mm}
We will examine this list of assertions in more detail in Section 2. The first two statements are a direct consequence of the fact that the soft hair is encoded in the diffeomorphism $f$ that relates the near horizon region to the asymptotic Minkowski space-time ${\cal I}$, as indicated in Figure~\ref{fig:softhair}. Due to its non-local nature, soft hair can only be measured by an infalling observer that travels from the asymptotic region to the horizon.

Consider a (coherent or incoherent) superposition of two classical global black hole space-times with identical near horizon geometries but with different soft hair. An observer falling into this state uses gravitationally dressed observables, that are anchored in the asymptotic space-time and thereby sensitive to the diffeomorphism $f$. While the observer experiences the same local black hole space-time in both parts of the state, her physical operators ${\cal O}_{\nspc {f}}$ are different. So we should promote $f$ to a quantum operator $\hat{f}$.

The soft hair dependence of physical operators manifests itself via the commutator between a late-time observable ${\cal O}_{\nspc \hat{f}}$ that measures a Hawking mode and an early-time operator ${\cal O}_{\smpc b}$ that creates an infalling mode, c.f. \cite{Kiem:1995iy}. The backreaction due to the infalling particle affects the soft hair $\hat{f}$, which in turn modifies the late observable ${\cal O}_{\nspc \hat{f}}$. This is the black hole butterfly effect~\cite{Shenker:2013pqa,Shenker:2013yza}. The expectation value of the (commutator)${}^2$ of ${\cal O}_{\nspc \hat{f}}$ and ${\cal O}_{\smpc b}$ defines an out of time ordered correlator that during the black hole scrambling period grows exponentially with time. On the gravity side, the Lyapunov growth follows from the shockwave interaction near the horizon \cite{dray1985}; in the microscopic theory, it is the manifestation of maximal chaotic quantum dynamics. The Lyapunov growth induces a quantum-classical transition through which soft hair becomes part of the classical space-time.

\vspace{-2.5mm}

\newsubsection{~Soft hair labels code subspace}

A common characteristic of known systems with maximal many body quantum chaos, such as the SYK model~\cite{KitaevTalk1,KitaevTalk2,Maldacena:2016hyu,Almheiri:2014cka,Maldacena:2016upp,Jensen:2016pah,Engelsoy:2016xyb}, is that their collective dynamics gives rise to an emergent Goldstone mode in the form of a diffeomorphism $f$. In the holographic bulk dual, the diffeomorphism $f$ represents the soft hair degree of freedom. In the following, we will assume that the four-dimensional black holes with HPS soft hair can also be given a holographic dual description with similar characteristics. We will continue to refer to this putative dual quantum system as `the holographic CFT'.  It would be interesting to understand how these statements fit into the emerging Celestial CFT dictionary~\cite{Pasterski:2016qvg,Pasterski:2017kqt,Pasterski:2017ylz}.

Expectation values in the  holographic CFT involve averaging over the full phase space of soft hair variables $f$. Schematically
\bea
\label{rhoeff}
\la  {\cal O}_{\hat f}\, ...\; {\cal O}_{\hat f} \ra_{\psi}\, = \;  \int [df]\, \rho_\psi(f) \; \la {\cal O}_f\, ...\; {\cal O}_f \ra_f
\eea
with $\rho_\psi(f)$ some probability distribution that depends on the CFT state $\psi$ and $\la {\cal O}_{\hat f}\, ...\; {\cal O}_{\hat f} \ra_f$ denotes the expectation value in an approximate $\hat{f}$ eigenstate with given soft hair $f$. Note that these approximate eigenstates themselves can still be mixed states with non-zero entropy.

We will distinguish two types of quantum states $\psi$:

\vspace{-2mm}

\begin{itemize}
\item{\it Soft focus states}: these are superpositions of states with different soft hair quantum numbers. Such states do not describe a fixed semi-classical global black hole space-time, but an incoherent superposition of global black hole space-times. The soft hair of soft focus states is in an undetermined state and carries a large amount of entropy, denoted by $S_{so\! \spc f\!\spc t}$. In the notation of \eqref{rhoeff}, we have
\bea
\label{ssoft}
S_{so\! \spc f\!\spc t} \, = \; - \int [df] \, \rho_\psi(f) \log \rho_\psi(f).
\eea
\item{\it Sharp focus states}: these are approximate eigenstates with given soft hair quantum numbers $f$ that describe a given semi-classical global black hole space-time. Sharp focus states carry their soft hair in well-determined state, and therefore have substantially less entropy than the soft focus states. We will call the Hilbert subspace ${\cal H}_f$ spanned by all states with the same soft-hair quantum numbers a code subspace. The microscopic entropy of the code subspace is denoted by $S_{code}$. 
\end{itemize}

The thermal CFT state is a soft focus state. It saturates the Bekenstein-Hawking entropy bound and represents a superposition of many global black hole space-times with different soft hair quantum numbers. As a maximal entropy state, it also maximizes the entropy $S_{so\! \spc f\!\spc t}$ carried by the soft modes. Since $S_{so\! \spc f\!\spc t}$ contributes to the total Bekenstein-Hawking entropy, we may write
\bea\label{ssplit}
{S_{B\nspc H} \, = \, S_{so\! \spc f\!\spc t} \, +\, S_{code}}
\eea
where $S_{code}$ accounts for the microscopic entropy of the code subspace, defined as the Hilbert subspace with fixed soft hair. Stated more microscopically, we can decompose the Hilbert space of the black hole into a direct sum of soft hair eigen sectors:
   $\mathcal{H}_{\nspc B\nspc H} \, =\, {\bigoplus}_f~ \mathcal{H}_{f}$. 
Assuming that the subspaces ${\cal H}_f$ with given soft hair $f$ are all isomorphic to some fixed abstract code subspace ${\cal H}_{code}$ with dimension $e^{S_{code}}$, we can construct all black hole states via an isometric embedding $\mathcal{H}_{code} \otimes \mathcal{H}_{so\! \spc f\!\spc t}  \ \hookrightarrow \    \mathcal{H}_{\nspc B\nspc H}$, where $\mathcal{H}_{so\! \spc f\!\spc t}$ denotes the abstract Hilbert space of the soft hair modes, spanned by the soft hair eigenstates $|f\ra_{\! so\! \spc f\!\spc t}$. 

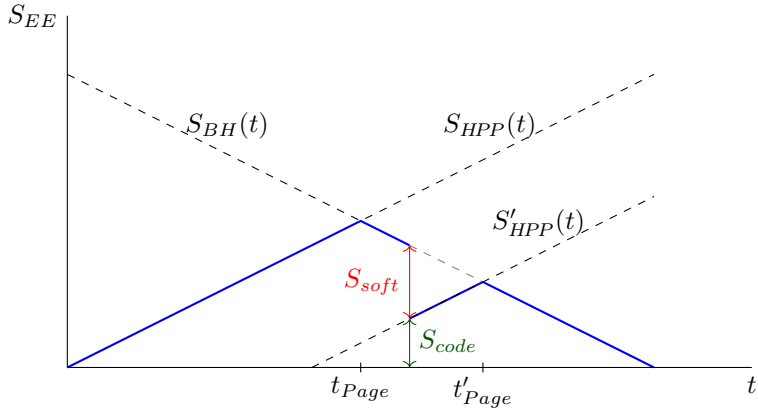
\begin{figure}[t]
\centering
\begin{tikzpicture}[scale=1.3]
\draw[thick, blue] (-3,0)--(0,1.5)--(.5,1.25);
\draw[thin,dashed,gray] (0.5,1.25) --(1.25,.875);
\draw[thick,blue] (1.25,.875) -- (3,0);
\draw[dashed] (-3,3)-- node[above]{$~~~S_{BH}(t)$}(0,1.5)--node[above]{$S_{H\! P
\!\spc P}(t)~~~~$}(3,3);
\draw (-3,0)--(-3,3.6)node[ left]{$S_{EE}$};
\draw (-3,0) --(0,0)node[below]{$t_{Page}$}-- (1.25,0)node[below]{$t'_{Page}$} -- (4,0) node[below ]{$t$};
\draw (0,-.05)--(0,.05);
\draw (1.25,-.05)--(1.25,.05);
\draw[red,<->] (0.5,1.25)--(0.5,0.875)node[left]{$S_{so\nspc f\nspc t}$}--(0.5,0.5);
\draw[color={rgb:green,2; black,4},<->] (0.5,0)--(0.5,0.275)node[right]{$S_{code}$}--(0.5,0.5);
\draw[thick,blue](0.5,0.5)--(1.25,0.875);
\draw[dashed] (-.5,0)--(2,1.25)node[above]{$S'_{H\! P
\!\spc P}(t)~~~~$}-- (3,1.75);
\end{tikzpicture}
\caption{\addtolength{\baselineskip}{-.5mm}Page curve~\cite{Page:1993df,Page:1993up} describing the evolution of entanglement entropy of the black hole with an infalling observer.  The Bekenstein bound, $S_{EE} \leq S_{BH}$ implies there must be a transition preventing the entanglement entropy from continuing to grow at late times, as would be predicted by Hawking pair production, $S_{H\! P\!\spc P}(t)$. The red segment indicates change in entropy when an infalling observer measures soft hair while crossing the horizon. After the measurement, the black hole entropy again follows the upward slope $S'_{H\! P\!\spc P}(t)$ prescribed by the Hawking process.}
\vspace{-3mm}
\label{fig:qtprotocol}
\end{figure}

\newsubsection{~Measurement of soft hair}

As we will argue, the infalling observer is able to measure the soft hair $f$, and thus naturally uses the sharp focus perspective. This has direct repercussions for the firewall argument. Depending on whether we adopt a Copenhagen or many worlds interpretation, the observer will (i) either project the black hole state onto an approximate soft hair eigenstate or (ii) become entangled with the soft hair. Crucially, since the soft hair is a property of the exterior geometry, she can do this while she is still outside the black hole. In both interpretations, the observer is able to evade the firewall obstruction. 

In case (i), the projection onto a given soft sector $|f\rangle_{\! so\! \spc f\!\spc t}$ reduces the entropy on the black hole space-time from $S_{B\nspc H}$ to $S_{code}$. The projection transforms the post-Page time black hole into a pre-Page time black hole with a density matrix that is contained within a single code subspace $\mathcal{H}_{code}$, labeled by the measured soft hair $f$. As we will review in section 5, this allows for the reconstruction of the Hawking partners inside the black hole via the Petz map~\cite{petzmap,work-in-progress}. 

In case (ii), the observer is able to avoid the firewall by the use of a `state dependent' set of observables ${\cal O}_{\nspc f}$ that depend on the soft hair quantum number $f$. More generally, her state-independent observables ${\cal O}$ take the form of a sum over an orthogonal set of soft hair quantum numbers 
\bea 
\label{ofsum}
{\cal O} \is \sum_f\, {\cal O}_f \, 
\otimes
|f\rangle_{O\nspc}\langle f|
\eea
where $|f\rangle_{O\nspc}$ denotes the state of the observer moving in the black hole space-time with given soft hair $f$. The $f$ quantum numbers of the space-time and the observer are correlated, since the observer can perform a measurement: the soft hair is classical information that can be measured and shared.

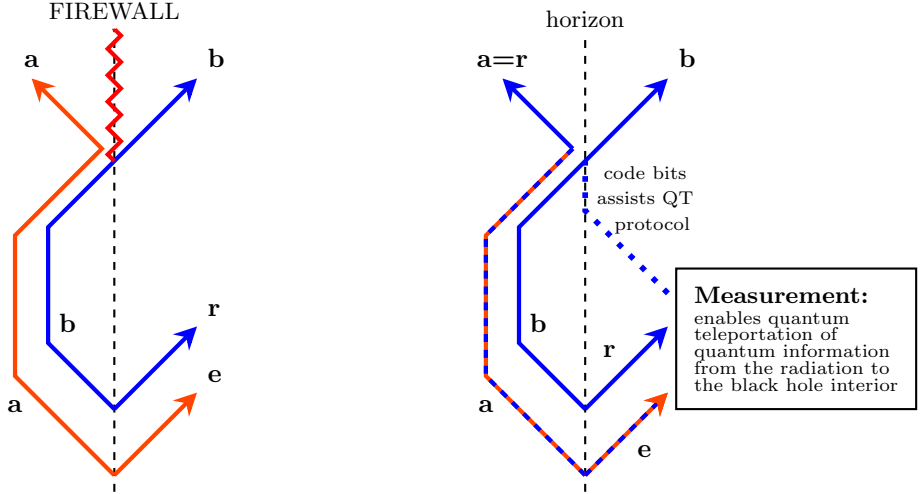
\begin{figure}[t]
\vspace{-2mm}
\centering
${}$\hspace{1cm}~\begin{tikzpicture}[scale=1.1]
\definecolor{amber}{rgb}{1.0, 0.49, 0.0};
\definecolor{burgundy}{rgb}{1, 0.0, 0.0};
\definecolor{orange-red}{rgb}{1.0, 0.27, 0.0}
\draw[dashed,thick] (6,-.2)--(6,4.9+.5) node[above]{\footnotesize FIREWALL};
\draw[color=blue,ultra thick,-{Stealth[length=3mm,width=3mm]}] (6,.8) -- (5.2,1.6) node[above right,color=black]{\bf b} --(5.2,2.5+.5)--(7,4.3+.5) node[above right,color=black]{\bf b};
\draw[color=blue,ultra thick,-{Stealth[length=3mm,width=3mm]}] (6,.8) --  (7,1.8) node[above right, color=black]{\bf r};
\draw[color=orange-red,ultra thick,-{Stealth[length=3mm,width=3mm]}] (6,0) -- (4.8,1.2) node[below,color=black,bag,font=\linespread{0.8}\selectfont]{\\\bf a} --(4.8,2.4+.5) -- (5.85,3.45+.5) -- (5, 4.3+.5)  node[above,color=black]{\bf a};
\draw[color=orange-red,ultra thick,-{Stealth[length=3mm,width=3mm]}] (6,0) --  (7,1) node[above right, color=black]{\bf e};
\draw[decorate,ultra thick,color=burgundy,decoration=zigzag](6,3.3+.5)--(6,4.9+.5);
\end{tikzpicture}\hspace{3cm}
\begin{tikzpicture}[scale=1.1]
\definecolor{amber}{rgb}{1.0, 0.49, 0.0};
\definecolor{burgundy}{rgb}{0.5, 0.0, 0.13};
\definecolor{orange-red}{rgb}{1.0, 0.27, 0.0}
\draw[dashed,thick] (6+1,-.2)--(6+1,4.8+.5)node[above]{\small horizon};
\draw[color=blue,ultra thick,-{Stealth[length=3mm,width=3mm]}] (6+1,.8) -- (5.2+1,1.6) node[above right,color=black]{\bf b} --(5.2+1,2.5+.5)--(7+1,4.3+.5) node[above right,color=black]{\bf b};
\draw[color=blue,line width=.75mm,loosely dotted] (7+1,2.2)--  (6+1,3.2) node[right, color=black,bag,font=\linespread{0.8}\selectfont]{\scriptsize code bits \\\scriptsize assists QT \\ \scriptsize ~~protocol \\} --(6+1,3.8) ;
\draw[color=blue,ultra thick,-{Stealth[length=3mm,width=3mm]}] (6+1,.8) -- node[above left, color=black]{\bf r}  (7+1,1.8) ;
\draw[color=orange-red,ultra thick] (6+1,0) -- (4.8+1,1.2) node[below,color=black,bag,font=\linespread{0.8}\selectfont]{\\\bf a} --(4.8+1,2.4+.5) -- (5.85+1,3.45+.5);
\draw[color=blue,ultra thick,-{Stealth[length=3mm,width=3mm]}]  (5.85+1,3.45+.5) -- (5+1, 4.3+.5)  node[above,color=black]{\bf a=r};
\draw[color=blue,ultra thick,dashed] (6+1,0) -- (4.8+1,1.2)  --(4.8+1,2.4+.5) -- (5.85+1,3.45+.5);
\draw[color=orange-red,ultra thick,-{Stealth[length=3mm,width=3mm]}] (6+1,0) -- node[below right, color=black]{\bf e} (7+1,1) ;
\draw[color=blue,dashed,ultra thick,-{Stealth[color=orange-red,length=3mm,width=3mm]}] (6+1,0) --  (7+1,1);
\draw[fill=white,thick,bag,align=left,font=\scriptsize\linespread{0.8}\selectfont] (7+1.1,0.8) rectangle node[]{\bf \small Measurement: \\[.5mm]  enables quantum \\ teleportation of\\
quantum information \\ from the radiation to \\ the black hole interior  } (7+4,2.5);
\end{tikzpicture}\\
\caption{\small \addtolength{\baselineskip}{-.5mm}Schematic depiction of the firewall argument and its proposed resolution via a quantum teleportation protocol. Left: quanta $e$ and $r$ escape the black hole thanks to entanglement with Hawking partners $a$ and $b$. After the Page time, when the black hole is in a maximally mixed state, $a$ and $b$ can not escape via the usual Hawking process: 
when $b$ leaves, it can not be entangled with any interior mode $a$. The horizon turns into a firewall. Right: as Hawking quanta propagate through space-time, they
entangle with their environment, they decohere and produce classical information encoded in the soft hair degrees of freedom. The classical information gets causally dispersed in the form of `code bits' that specify the code subspace. The Hawking partner $b$ can escape the black hole via the usual Hawking process, with help from a quantum teleportation protocol enabled by the code bits.}
\label{QTprotocol}
\vspace{-2mm}
\end{figure}

Both proposed resolutions of the firewall are not new. The new element in our story is that we propose a concrete physical justification for the code space projection, or code subspace dependence, via the outside measurement of the classical HPS soft hair degrees of freedom. It is worth noting that our proposed resolution of the firewall paradox makes use of an effective quantum teleportation protocol, as indicated in Figure~\ref{QTprotocol}.

\smallskip

The remainder of this paper is organized as follows. In section 2 we present a more detailed description of the physical properties 1 through 7 of the black hole soft hair listed above. Our discussion here will borrow from HPS and follow up work. The main new ingredients are the characterization of soft hair in terms of a transition functions, and the demonstration of its Lyapunov behavior.  In Section 3 we introduce an observer dependent notion of the firewall, and show that the infalling observer never encounters her own firewall before reaching the singularity. In Section 4, we present the construction of the interior operators\cite{Verlinde:2012cy}, based on technology borrowed from approximate error correction  and the Petz map\cite{Cotler:2017erl}. We end with some concluding comments in section 5. In the Appendix we describe the soft hair phase space of the black hole.

\section{Measuring Black Hole Soft Hair}

\vspace{-2mm}

In this section we examine assertions 1-7 about the physical properties of soft hair that underlie our proposed resolution of the firewall paradox. The basic physical statement that will organize our discussion is that the black hole soft hair degrees of freedom are encoded in the transition function between different coordinate regions of the global black hole space-time. Specifically, we can divide an eternal black hole space-time into four different regions, corresponding to past and future null infinity ${\cal I}^\pm$ and the past and future horizon ${\cal H}^\pm$. In the following, we will combine the past and future horizon into a single near horizon region ${\cal H}$.\footnote{Alternatively, we can consider a collapsing black hole geometry with only a future horizon. Here we prefer to work with an eternal black hole geometry with fixed boundary conditions at the past horizon. In either case, only the future horizon will carry soft hair. }

\subsection{Soft hair as a transition function}
\vspace{-2mm}

In both asymptotic regions and the near horizon region, the soft degrees of freedom manifest themselves through the presence of an asymptotic symmetry group, the BMS group~\cite{Bondi:1962px,Sachs:1962wk}. At the linearized level, this symmetry group acts on the space-time metric via infinitesimal diffeomorphisms, called supertranslations. In both regions, the metric with supertranslation degrees of freedom looks like an infinitesimal variation of the unperturbed black hole metric
\bea
\label{reparam}
ds^2 = (g_{ab} + h_{ab}) dx^a dx^b, \qquad \qquad h_{ab} ={\cal L}_f g_{ab}
\eea
with ${\cal L}_f g_{ab}$ the Lie derivative of the background metric $g_{ab}$ with respect to a suitable vector field $\zeta_f$ associated with the supertranslation parameter $f$. Hence locally, the soft hair degree of freedom can be transformed away via a diffeomorphism: we can choose local coordinates so that the asymptotic metric looks like ordinary Minkowski space and the near horizon region looks like a standard black hole metric with no hair. The soft hair is then encoded in the transition function between the two regions. 

To write the metric in both regions, we will choose light cone coordinates $(u,v) = (t-r^*,t+r^*)$ with $r^*=r+2M\ln|{r}/{2M}-1|$ in the asymptotic region and Kruskal-Szekeres coordinates $(U,V)$ in the near horizon region. The asymptotic metric takes the form
\bea
\label{lcmetric}
ds^2_{\; \mbox{$|$}{\rm near} \ {\cal I}^\pm} \is  - \Lambda \spc du dv + r^2\gamma_{AB}dz^Ad z^B, \qquad \qquad \Lambda \equiv 1 - \frac{2M}{r}
\eea
and reduces to standard Minkowski space for large $r$. The metric in the near horizon region takes the standard Kruskal form
\bea
\label{kruskalm}
ds^2_{\; \mbox{$|$}{\rm near} \ {\cal H} } \is - F\spc dU dV + r^2\gamma_{AB}dZ^Ad Z^B, \qquad \qquad F \equiv \frac{2M}{r}\spc e^{-r/2M}
\eea
with $\gamma_{AB}$ the metric on $S^2$. The metric \eqref{kruskalm} is appropriate for describing the black hole region as it reduces to the standard Minkowski space metric (up to a factor of $e$) near the bifurcate horizon $U=V=0$. 

In the absence of soft hair, the two coordinate systems are patched together at some small macroscopic distance outside the black hole, via the coordinate transformation $U/{4M}\! =\! -e^{-u/4M}$, ${V}/{4M}\! =\! e^{v/4M}$, $Z^A  = \spc z^A$.
The linearized soft hair degrees of freedom will modify this transition function into a relation of the form
\bea
\label{trans2}
\frac U{4M}\spc =\spc -e^{-u/4M} + \zeta_f^U ,~~&&~\frac{V}{4M}\spc =\spc e^{v/4M} + \zeta_f^V,\qquad \ \ Z^A  = \spc z^A + \zeta_f^A\\ \nonumber
\eea
where $\zeta_f^U$, $\zeta_f^V$  and $\zeta_f^A$ denote components of the vector field $\zeta_f$ associated with the infinitesimal supertranslation labeled by $f$. The space-time with the two coordinate regions is indicated in Fig~4. Thanks to the introduction of the transition function, we can stipulate that the near horizon $(U,V)$ geometry still takes the standard no hair form \eqref{kruskalm}, while the metric in the $(u,v)$ region carries non-zero supertranslation hair: it satisfies the boundary condition that at the transition region, it takes the form in equation \eqref{reparam} with $g_{ab}$ as in \eqref{lcmetric}.  We will determine the explicit form of $\zeta_f$ in the next subsection using the results of HPS.

\subsection{Schwarzschild supertranslations}

\vspace{-2mm}

In HPS, the soft hair degrees of freedom of the black hole are identified through the residual supertranslation invariance of the Schwarzschild metric in the advanced Bondi gauge.  The standard Schwarzschild metric in the advanced Bondi coordinates reads~\cite{Hawking:2016sgy}  
\bea
  \label{bhmetric}
  ds^2=-\Lambda dv^2+2 dvdr+r^2\gamma_{AB}dz^Ad z^B
  \eea
with $\Lambda$ defined in \eqref{lcmetric}. Following ~\cite{Hawking:2016sgy}, we now consider BMS$^-$ supertranslations which preserve Bondi gauge and the standard falloffs at large $r$. The infinitesimal supertranslations are generated by the vector field $\zeta_f$ of the form
\be\label{ssy} \zeta_{{}_{f}}= {f} \smpc\p_v-\half D^2f\smpc \p_r +{1 \over r}D^Af\smpc \p_A .
\ee 
Here the function $f$ only depends on the transverse $S^2$ coordinates $\zet^A$. The explicit form of the supertranslated Schwarzschild geometry is
 \bea\label{sf} ds^2\is -\Bigl(\Lambda -{MD^2f \over r^2}\Bigr)dv^2+2 dvdr- dvd\zet^AD_A(2\Lambda f+D^2f) \cr\cr &&~~~~~~+
\bigr(r^2\g_{AB}+2rD_AD_Bf-r\g_{AB}D^2f\bigr)d\zet^Ad\zet^B. \eea
This metric is of the form \eqref{reparam}. Note that the supertranslation shifts the horizon to $r = 2M + \frac 1 2 D^2 f$.

In the terminology of~\cite{Hawking:2016sgy},  the black hole described by \eqref{bhmetric} is said to have linearized supertranslation hair. We will make this statement more precise in the next subsection, where we discuss the combined action of the supertranslation generators on the three separate regions, the two asymptotic regions ${\cal I}^\pm$ and the near horizon region ${\cal H}$. Indeed, it is an important property of soft hair that it can only be specified by considering the global space-time geometry in all three regions.

Via the above mentioned coordinate transformation to the $(u,v)$ and $(U,V)$ coordinate systems, we can rewrite the supertranslation vector field as follows
\bea 
\label{newzeta}
\zeta_{{}_{f}}
\! \is \! e^{v/4M} f \p_V + \Bigl( {e^{-u/4M}} f + 
\gamma\, e^{-v/4M} \spc D^2f\Bigr)\p_U  +  {1 \over r}D^Af \p_A, \qquad \gamma \equiv 1/F.
\eea
Note that while we have changed our coordinate system, this is still using BMS$^-$ gauge fixing. This explains the asymmetry between the ingoing and outgoing lightcone coordinates. Inserting  \eqref{newzeta} into  \eqref{trans2}, we find that the transition function between the near horizon and asymptotic light cooordinates takes the form
\bea
\label{patch1}
\frac{{V}}{4M}  \is 
e^{v/4M}  \Bigl( 1+ \frac{f}{4M}\Bigr)\\[2mm]
\frac{{U}}{4M}  \is 
-\spc e^{-u/4M} \Bigl( 1- \frac{f}{4M} - \frac \gamma {4M} \spc e^{(u-v)/{4M}}\spc D^2 {f} \Bigr)\\[2.5mm]
{Z}^A \is z^A + \frac 1 r D^A f.
\eea
We can invert these relations as
\bea
\label{patch2}
{v}\is 4M \log\Bigl(\frac{V}{4M}\Bigr) - {f}\\[1.5mm]
\label{patchtwo} {u} \is - 4M \log\Bigl(-\frac{U}{4M}\Bigr)  -\spc {f} - \,  \gamma \spc e^{(u-v)/{4M}}\spc D^2 {f}\\[2mm]
z^A\!\! \is Z^A - \frac 1 r D^A f.
\eea
The $e^{(u-v)/{4M}}$ factor in the last term in \eqref{patchtwo} exhibits the exponential redshift effect near the black hole horizon.\footnote{Note that $\gamma = (2M/r) e^{r/2M}$ approaches a constant $\gamma=e=2.718...$ at the horizon.}  It enhances the influence of small horizon shifts on future light-like trajectories. This exponential effect was emphasized in earlier works~\cite{DRAY1985173,dray1985,Kiem:1995iy}, and more recently identified as a manifestation of Lyapunov behavior of the underlying microscopic dynamics.  
The important observation for our purpose is that the Lyapunov growth enforces a quantum-classical transition from an early state where part of the soft hair can be in a quantum superposition to a later state where the initial quantum mechanical soft hair has become part of the classical space-time. We will return to this quantum-classical transition later on.

\begin{figure}[t]
\centering
\begin{tikzpicture}[scale=.78]
\definecolor{darkgreen}{rgb}{.0, 0.5, .1};
\draw[fill=gray!20,opacity=.3,dashed] (0,6) .. controls (.4,4) and (.4,2) ..  (0,0);
\draw[red,opacity=.8, dashed] (0,6) .. controls (-.4,4) and (-.4,2) ..  (0,0);
\draw[blue] (0,3) node {${f}$};
\draw[fill=gray!20,dashed,opacity=.3] (0,6) .. controls (-.4,4) 
and (-.4,2) ..  (0,0);
\draw[color={rgb:green,5; black,3}, dashed] (0,6) .. controls (.4,4) and (.4,2) ..  (0,0);
\draw[thick] (0,6)node[above ] {\ $i^+$}--node[right] {\, $\mathcal{I}^+$} (3,3) node[right ] {$i^0$} node[left]{\mbox{$\textcolor{red}{\large ({u},{v},{z})}$\quad\ }}-- node[right] {\ $\mathcal{I}^-$\ } (0,0)node[below ] {$i^-$} -- node[left] {\ $\mathcal{H}^-$}  (-3,3)node[right]{\mbox{\quad $\textcolor{darkgreen}{\large ({U},{V},{Z})}$}} -- node[left] {\ $\mathcal{H}^+$}(0,6);
\draw [thick](-4,2) -- (-3,3) -- (-4,4); 
\draw [thick,decorate,decoration=snake] (0,6) -- (-4,6);
\draw [thick,decorate,decoration=snake] (0,0) -- (-4,0);
\end{tikzpicture}
\vspace{-2mm}
\caption{\addtolength{\baselineskip}{-.5mm} The soft hair degrees of freedom are encoded in the transition function $f$ between near horizon Kruskal coordinates $(U,V)$ and asymptotic $(u,v)$ coordinates.}
\label{fig:sigma2}
\end{figure}
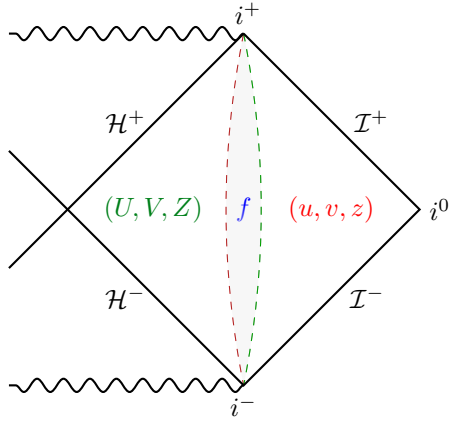
\vspace{-2mm}

\subsection{Asymptotic and horizon symmetries}
\vspace{-2mm}

The soft hair degrees of freedom locally look like gauge redundant coordinate transformations.  What makes them a true physical property of the black hole space-time? HPS present two arguments. The first is based on the fact that the BMS symmetries of future and past asymptotic infinity imply the existence of conserved supertranslation charges that commute with the full gravitational S-matrix~\cite{He:2014laa,Hawking:2016sgy}
\bea\label{eq:ward}
   \la out\ri Q^+{S}-{S} Q^-\li in\ra \is  0.
\eea
This equation holds locally on every point on the celestial sphere.
The reasoning is as follows. Canonical charges in gauge theory and gravity reduce to boundary terms.  If we pick a Cauchy slice for our space-time this boundary will be near spatial infinity $i^0$.  Choosing a slice that hugs future (past) null infinity in the early $u$ (late $v$) region, the charge can be evaluated in terms of gauge field or metric components near $\mathcal{I}^+_-$ ($\mathcal{I}^-_+$).  An antipodal matching condition~\cite{Strominger:2013jfa} follows from CPT invariance and thus equates charges on Cauchy slices that capture incoming and outgoing radiation, respectively.
In the vacuum case -- with only massless matter and no black holes --  future and past null infinity form Cauchy slices.  Equating the charges for these two slices provides a non-trivial constraint on low energy scattering.  This is because consistency of the constraint equations on each slice separately, combined with the matching near $i^0$, gives a non-trivial relation between hard and soft modes.  Classically these are called memory effects~\cite{1974SvA,1987Natur,Strominger:2014pwa}.  Within S-matrix insertions these are soft theorems~\cite{Weinberg:1965nx}.

In the presence of a black hole, null infinity is no longer a Cauchy slice and we must include the horizon.  The constraint equations still hold.  When performing a large gauge transformation, this coordinate change must be propagated through the bulk and will act on the horizon.  Let ${\cal S}$ denote the unitary quantum evolution operator associated with the exterior $(u,v)$ coordinate region indicated in Figure 2. It maps initial states $\li in\ra$ defined at past null infinity ${\cal I}^-$ to final states at future null infinity ${\cal I}^+$ and the horizon region~${\cal H}$. 
Following HPS, we can now express the conservation of BMS charge via a relation of the form\footnote{To write this equation in more explicit form, 
let $|r \rangle_{\! R}$ denote a complete bases of asymptotic radiation states at ${\cal I}^+$ and 
$| b \rangle_{\! B}$ a complete bases of black hole states and let 
$$
{\cal S}\li {in} \ra \,=\, \sum_{r,b}\;  A_{rb} \, \li r \ra_{\! R} \, \li b \ra_{\! B}
$$
with $A_{rb}$ some set of amplitudes. The conservation equation \eqref{qsplit} of supertranslation charge then reads
$$
 {\cal S} \,  Q^{{\cal I}^+}_{f_1} \li in \ra\, =\, \sum_{r,b } A_{rb}\, \Bigl(Q_{f_2}^{{\cal I}^+}\li r \ra_{\! R} \otimes \li b\ra_{\! B} \, + \, \li r\ra_{\! R} \otimes Q_{f_3}^{{\cal H}^+} \li b\ra_{\! B}\Bigr).
$$
}
\bea
\label{qsplit}
{\cal S} \, Q^{{\cal I}^-}_{f_1}\! \is \bigl(\spc Q^{{\cal I}^+}_{f_ 2} + Q^{{\cal H}}_{f_3} \bigr)\,  {\cal S}.
\eea
Here $Q^{{\cal I}^-}_{f_1}\!\!,$ $Q^{{\cal I}^+}_{f_ 2}$ and $Q^{{\cal H}}_{f_3}$ denote supertranslation generators acting on the corresponding regions. Their supertranslation parameters $f_1$, $f_2$ and $f_3$ are linearly related via propagation of the linearized coordinate change through the bulk. Otherwise, the parameters can be chosen freely. Equation \eqref{qsplit} thus amounts to a local relation on the celestial sphere. We will give a more explicit form of the supertranslation generators in the next subsection.

The conservation relation \eqref{qsplit} implies that there is a residual gauge invariance, that acts by simultaneously shifting the soft hair degrees of freedom in the three regions. The true global gauge invariant notion of the soft hair of the black hole is defined as the gauge equivalence class of the three local soft hair modes, modulo the overall simultaneous shift. This gauge invariance allows one to gauge away the local soft hair at any one of the three regions. A convenient gauge choice is to prescribe  that the geometry in the far past takes the standard form, without any soft hair. The black hole soft hair is then encoded in the patching function \eqref{patch1}-\eqref{patch2} that connects the outside $(u,v)$ region to the near horizon $(U,V)$ region. 

Since $f$ can be some arbitrary function of the angular coordinates $z^A$, the number of soft hair degrees of freedom scales as the area of the black hole horizon measured in units of the UV cut-off. So assuming one can place this UV cut-off close the Planck scale, the entropy encoded in the soft hair could add up to a finite fraction of the Bekenstein-Hawking entropy. The key observation of HPS is that the soft hair entropy $S_{so\!f\!t}$ should a) be thought of as part of the black hole entropy, and b) is associated with non-local but a priori observable features of the external  space-time geometry. 
Via the above characterization of the HPS soft hair degrees of freedom, we have given supporting evidence for properties 1,2 and 5 on our list -- that soft hair is inherently non-local and thus locally invisible and can carry a large amount of entropy -- and have seen some hints in support of properties 3, 4 and 6.

\subsection{Implanting soft hair} 
\vspace{-2mm}
A second physical argument for the existence of black hole soft hair presented by HPS is that
supertranslation hair can be `implanted' by means of sending in a massless matter shockwave in from $\mathcal{I^-}$ along some light-like trajectory $v=v_0$.
Such a shockwave is described by the following stress tensor~\cite{Hawking:2016sgy} 
\bea
\label{tansatz}
    \hat{T}_{vv}=\left(\frac{\mu+\hat{T}}{4\pi r^2}+\frac{\hat{T}^{(1)}}{4\pi r^3}\right)\delta(v-v_0),~~&&~~\hat{T}_{vA}=\frac{\hat{T}_A}{4\pi r^2}\delta(v-v_0)
\eea
where $\hat{T}$, $\hat{T}^{(1)}$ and  $\hat{T}_A$ depend on the transverse coordinates $z^A$ only. The ansatz \eqref{tansatz} satisfies $\nabla^\mu T_{\mu\nu}=0$ throughout the bulk by means of the following relations between the subleading and leading-in-$r$ modes:
\bea\label{cons}
    (D^2+2)\hat{T}^{(1)}=-6M\hat{T},\ \  & & \ \ D^A\hat{T}_{A}=\hat{T}^{(1)}
\eea
with  $D_A$ the covariant derivative along the sphere and $D^2 = D^AD_A$. 

When inserted into the Einstein equation, this lightlike matter pulse induces a shift in the transverse components of the metric on the horizon.  Parametrizing the general solution to the conservation equation~\eqref{cons} in the suggestive form 
\bea\label{eq:tmunu}
  \hat{T} \! \is \! -\frac{1}{4}D^2(D^2+2){\spc\hat{f}}, \qquad \quad \hat{T}_{A}=\frac{3M}{2}D_A\spc\hat{f}, \qquad \quad \hat{T} = \frac{3M}{2} D^2 \spc\hat{f}
\eea
we find that the solution is diffeomorphic to Schwarzschild both before and after the shockwave. Assuming the initial black hole geometry has no supertranslation hair, the solution with the shockwave takes the form
\be\label{eq:shock1} h_{ab}=\Theta(v-v_0)\left(\cL_{\spc\hat{f}\spc}g_{ab}+{2\mu \over r}\delta_a^v\delta_b^v\right)\ee
with $h_{ab}$ the deviation of the unperturbed black hole metric~\eqref{bhmetric}.  We thus see that the shockwave is a domain wall interpolating between two 
BMS inequivalent Schwarzschild vacua, whose mass parameters differ by $\mu$. At the quantum level, this describes a coherent state of soft gravitons. 
Note that the $\ell=\{0,1\}$ modes of $\hat{f}$ are in the kernel of $D^2(D^2+2)$. This is related to the fact that gravity does not produce dipole radiation.  

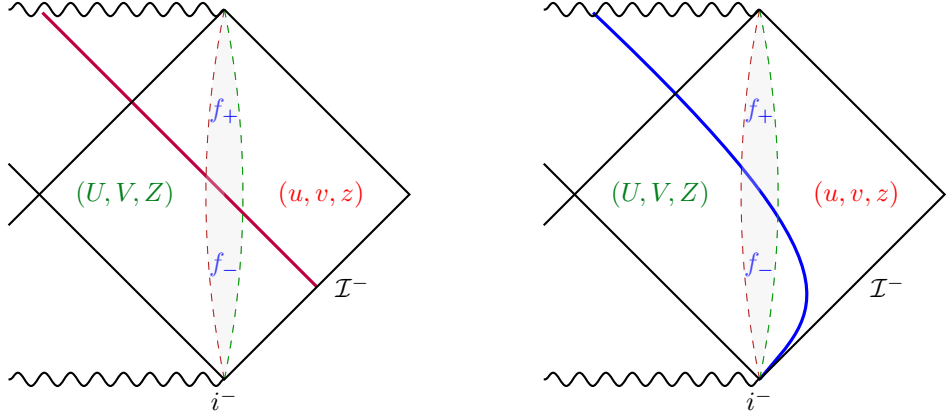
\begin{figure}[t]
\centering
\begin{tikzpicture}[scale=.82]
\definecolor{darkgreen}{rgb}{.0, 0.5, .1};
\draw[fill=gray!20,opacity=.3,dashed] (0,6) .. controls (.4,4) and (.4,2) ..  (0,0);
\draw[red,opacity=.8, dashed] (0,6) .. controls (-.4,4) and (-.4,2) ..  (0,0);
\draw[very thick,purple] (1.5,1.5) -- (-2.95,5.95);
\draw[blue] (0,4) node[above]{${f}_+$};
\draw[blue] (0,1.5) node[above]{${f}_-$};
\draw[fill=gray!20,dashed,opacity=.3] (0,6) .. controls (-.4,4) 
and (-.4,2) ..  (0,0);
\draw[color={rgb:green,5; black,3}, dashed] (0,6) .. controls (.4,4) and (.4,2) ..  (0,0);
\draw[thick] (0,6)
-- (3,3) 
node[left]{\mbox{$\textcolor{red}{\large ({u},{v},{z})}$\quad\ }}-- node[right] {\ $\mathcal{I}^-$\ } 
(0,0) node[below ] {$i^-$}   
-- (-3,3) node[right]{\mbox{\quad $\textcolor{darkgreen}{\large ({U},V,{Z})}$}}-- (0,6);
\draw[thick] (-3.5,2.5) -- (-3,3) -- (-3.5,3.5); 
\draw [thick,decorate,decoration=snake] (0,6) -- (-3.5,6);
\draw [thick,decorate,decoration=snake] (0,0) -- (-3.5,0);
\end{tikzpicture}
~~~~~~~~~~~~~~\begin{tikzpicture}[scale=.82]
\definecolor{darkgreen}{rgb}{.0, 0.5, .1};
\draw[fill=gray!20,opacity=.3,dashed] (0,6) .. controls (.4,4) and (.4,2) ..  (0,0);
\draw[red,opacity=.8, dashed] (0,6) .. controls (-.4,4) and (-.4,2) ..  (0,0);
\draw[very thick,blue] (0,0) .. controls (.9,1.1) and (1.9,1.6) ..  (-2.7,5.95);
\draw[blue] (0,4) node[above]{${f}_+$};
\draw[blue] (0,1.5) node[above]{${f}_-$};
\draw[fill=gray!20,dashed,opacity=.3] (0,6) .. controls (-.4,4) 
and (-.4,2) ..  (0,0);
\draw[color={rgb:green,5; black,3}, dashed] (0,6) .. controls (.4,4) and (.4,2) ..  (0,0);
\draw[thick] (0,6)
-- (3,3) 
node[left]{\mbox{$\textcolor{red}{\large ({u},{v},{z})}$\quad\ }}-- node[right] {\ $\mathcal{I}^-$\ } 
(0,0) node[below ] {$i^-$} 
-- (-3,3) node[right]{\mbox{\quad $\textcolor{darkgreen}{\large ({U},V,{Z})}$}}-- (0,6);
\draw[thick] (-3.5,2.5) -- (-3,3) -- (-3.5,3.5); 
\draw [thick, decorate,decoration=snake] (0,6) -- (-3.5,6);
\draw [thick, decorate,decoration=snake] (0,0) -- (-3.5,0);
\end{tikzpicture}
\vspace{-1mm}
\caption{\addtolength{\baselineskip}{-.5mm} Infalling matter shockwaves produce a discontinuity $\hat{f} = f_+-f_-$ in the soft hair mode encoded in the transition function between the two coordinate regions. For a lightlike matter pulse (left) the shockwave starts at past null infinity ${\cal I}^-$, for a massive matter particle (right) the trajectory starts at past infinity $i^-$.  }
\label{fig:sigma1}
\vspace{-1mm}
\end{figure}

The above solution is a slight generalization of the Dray-'t Hooft shockwave solution~\cite{dray1985,Kiem:1995iy}.  
In~\cite{DRAY1985173} and thus equation (1) of~\cite{dray1985} the metric for a shockwave moving along the past horizon is described.  At the horizon, equations \eqref{tansatz}-\eqref{eq:tmunu} reduce to
\bea\label{eq:tH}
    \hat{T}_{vv}\!\is\!\frac{1}{16\pi M^2}\left(\mu-\frac{1}{4}D^2(D^2-1)\spc\hat{f}\right)\delta(v-v_0),~~~~~\hat{T}_{vA}\, =\ \frac{3}{32\pi M}D_A\spc\hat{f}\,\delta(v-v_0).
\eea
Note that the $(D^2-1)$ factor is now negative definite and therefore invertible.
 
The presence of these shockwave solutions shows that soft hair degrees are real and that they can be changed by sending in matter into the black holes. Combined with the observation that the effect of supertranslation hair on future trajectories exhibits Lyapunov growth, this establishes property 4, that soft hair is an exponentially sensitive probe of infalling matter. The Lyapunov behavior manifests itself via out of time ordered correlation functions, or equivalently, by considering the (commutator)$^2$ of late time and early time observables. The imprint of the Dray-'t Hooft shockwave dynamics on such commutators was first considered in \cite{Kiem:1995iy}.

\subsection{Soft hair phase space}
\vspace{-2mm}

To understand the extent to which soft hair is a classical property of the external space-time geometry, it is important to map out the phase space structure of soft hair. In particular, we would like to find the canonical conjugate variable to the soft hair Goldstone mode $\hat{f}(z)$, or equivalently, the soft part of the supertranslation generator $Q_S(z)$ localized at a transverse point $z$, defined through the commutation relation\footnote{From here on we will use the shorthand $z$ to represent the point with coordinates $z^A=(z,\bz)$.  Functions of $z$ need not be holomorphic.}
\bea
\label{qcom}
\bigl[Q_S(z), \hat{f}(z')\bigr]\is 
 i\delta^{(2)}(z\!\spc -\!\spc z').
\eea
Here the hat notation indicates that $\hat{f}$ is now promoted to a quantum operator.

As discussed above, we can shift the value of the Goldstone mode $\hat{f}(z)$ by sending in a matter pulse through the horizon. To capture this potential time dependence, we must promote $f$ to a function of $z$ and $v$, the lightcone coordinate along the horizon. Similarly, we can write the soft charge $Q_S(z)$ as an integral of a charge current along the horizon
\bea
Q_S(z) \! \is \! \int_{-\infty}^\infty\! \!\! dv \, \hat{q}_S(v,z). 
\eea
By definition, the current $q_S(v,z)$ is the canonical conjugate variable of $\hat{f}(v,z)$
\bea
\label{cancom}
\bigl[\hat{q}_S(v,z), \hat{f}(v',z')\bigr]\is  i \delta(v\!\spc -\!\spc v')\,
\delta^{(2)}(z\!\spc -\!\spc z').
\eea
We would like to find an explicit expression for $\hat{q}_S(v,z)$.  HPS have already done this work for us.

The soft part of the horizon charge can be expressed in terms of the linearized metric perturbation $h_{ab}$. It is sufficient to first focus on the perturbation $h_{AB}$ of the transverse metric $\gamma_{AB}$. In the Bondi gauge, $h_{AB}$ is traceless. The analogue of its canonical momentum dual to $h_{AB}$ is the traceless shear tensor
\bea
\label{shear}
\sigma_{AB} = \frac 1 2  \partial_v h_{AB}.
\eea
In~\cite{He:2014laa}, it was shown that this mode has a nontrivial Dirac bracket with the supertranslation Goldstone mode at asymptotic infinity. A similar zero mode algebra appears at the horizon ${\cal H}$. 

In the horizon region, the metric perturbation $h_{AB}$ is parametrized by the soft hair mode $\hat{f}$ via
\bea
\label{hf}
h_{AB}|_{\mathcal{H}}\is 2M(2 D_A D_B-\gamma_{AB}D^2)\hat{f}.
\eea
Here $h_{AB}$ and $\hat{f}$ are both viewed as quantum mechanical operators.
Using the standard canonical form of the Einstein action, HPS derive the following commutation relations between the metric perturbation and the shear tensor at the horizon~\cite{Hawking:2016sgy}
\bea\label{bracket}
\bigl[ \sigma_{AB}(v,\zet), h_{CD}(v',\zet')\bigr] \is  32\pi i M^2 G_{\! ABCD} \, \delta(v\!\spc -\!\spc v')\, \delta^{(2)}(\zet\!\spc -\!\spc \zet')
\eea
with $G_{\! ABCD} = \gamma_{AC}\gamma_{BD}+ \gamma_{AD}\gamma_{BC}- \gamma_{AB}\gamma_{CD}$ the DeWitt metric on the space of 2D metrics $\gamma_{AB}$. 

Now consider the combination
\bea
\hat{q}_S \is \frac{1}{16\pi M} D^A D^B \sigma_{AB}.
\eea
Using \eqref{bracket}, we derive that
\bea
\bigl[\hat{q}_S(v,z),h_{AB}(v',z')\bigr] \is\, 2 iM(2 D_A D_B-\gamma_{AB}D^2)\,\delta(v\!\spc -\!\spc v') \, \delta^{(2)}(z\! -\! z'). 
\eea
Comparing with \eqref{hf} confirms that $q_S(v,z)$ is the sought after canonical conjugate to the soft mode $\hat{f}(v,z)$.
Moreover, using the expression \eqref{shear} for the shear tensor and the commutation relations for $D_A$ listed in Appendix A of \cite{Hawking:2016sgy}, we find that $q_S$ can be expressed in terms of the $v$ derivative of the Goldstone mode~as
\bea
\hat{q}_S  \is \frac{1}{ 16\pi}\, D^2(D^2+2) \, \partial_v \hat{f}. 
\eea
The canonical commutation relation \eqref{cancom} then integrates to 
\bea
\label{fcom}
\bigl[\spc \partial_v \hat{f}(v,\zet), \hat{f}(v',\zet')\spc\bigr] \is\! i\delta(v\!\spc -\!\spc v') \, \Omega(\zet,\zet')\\[3mm]
(D^2 +2) D^2  \Omega(\zet,\zet') \! \is  16\pi \,  \delta^{(2)}(\zet - \zet').
\eea
The explicit form of the Green's function $\Omega(z,z')$ is given in~\cite{He:2014laa}.
The commutation relation \eqref{fcom} is identical to the one derived for the asymptotic supertranslation modes at ${\cal I}^\pm$. However, there are still important physical differences between the horizon and asymptotic soft hair modes.

The non-trivial phase space structure of the soft hair modes puts limits on the extent to which they correspond to classically measurable properties of space-time. One of the  classical consequences of the soft hair dynamics and commutation relations is the gravitational memory effect.  The memory effect due to asymptotic soft dynamics at ${\cal I}^\pm$ is encoded in subtle infrared observables encapsulated by the Weinberg soft theorem for scattering amplitudes. The memory effect due to the horizon soft dynamics is more pronounced; it gets exponentially enhanced via the black hole butterfly effect.
The horizon memory mode is given by the integral of $\partial_v \hat{f}$ over the horizon
\bea
\int_{-\infty}^\infty\!\!\! dv\;\partial_v \hat{f}(v,z) \is \hat{f}_+(z)-\hat{f}_-(z)
\eea
where $\hat{f}_\pm(z)$ denote the value of the soft hair mode at the past and future end-point of ${\cal H}^+$. 
The horizon memory mode is related to the soft contribution to the total supertranslation generator via
\bea
Q_S(z) \is  \frac{1}{16\pi}D^2(D^2+2)\bigl(\hat{f}_+(z)-\hat{f}_-(z)).
\eea
Integrating \eqref{fcom}, we derive that the far future soft hair mode has the following commutator with the far past soft hair mode
\bea
\label{fpmcom}
\bigl[\hat{f}_+(\zet), \hat{f}_-(\zet')\bigr] \is\spc    i\Omega(\zet,\zet').
\eea

The commutator \eqref{fpmcom} is a quantum gravity effect. Reinstating units, we read off that the right-hand side contains a factor of the (Planck length)$^2$. So at first sight, it may look like the commutator has only microscopic consequences that can not be discerned by macroscopic observers. Note, however, that the soft mode $\hat{f}$ governs the location of the event horizon and, as emphasized earlier, any small shift of the horizon has exponentially growing consequences for light-like future trajectories. The resulting exponentially growing commutators are captured by the out-of-time ordered correlation functions. This black hole butterfly effect~\cite{Shenker:2013pqa,Shenker:2013yza} can lead to the breakdown of classical physics expectations, or conversely, catalyze the quantum-to-classical transition for other quantities~\cite{Zurek:1994wd,Zurek:1995jd}. 
We will make this exponential growth more explicit in the next subsections.

\subsection{Soft gravitational dressing} 
\vspace{-2mm}

Physical operators in a gravitational theory are diffeomorphism invariant combinations of operators made up from matter fields and gravitational fields. Keeping track only of the invariance under supertranslations, we need to require that physical operators must commute with the total supertranslation charge
\bea
\label{physcomm}
\bigl[Q_f, \mathcal{O}_{phys} \bigr] \is 0.
\eea 
Schematically, if we choose a suitable supertranslation eigen basis, we expect that these physical operators will factorize into a product $
\mathcal{O}_{phys} = \mathcal{O}\, \mathcal{W} $ of a matter operator times a suitable gravitational Wilson line observable.
We would like to find an explicit description of these physical observables in our setting.

Supertranslations act both on the gravitational soft Goldstone mode $\hat{f}$ and on the matter fields. The matter perturbations (which include gravitons) are localized in space-time, while the soft modes by definition are far-infrared degrees of freedom that can only be measured by $u$ or $v$ integrals over all of ${\cal I}^\pm$ or ${\cal H}$. Correspondingly, the supertranslation charges acting on the in and out states can be split into hard and soft pieces
   $Q^\pm=Q_S^\pm+Q_H^\pm.$
Here $Q_S^\pm$ induces inhomogeneous shifts in the geometric data at null infinity and, as discussed above, classically corresponds to a memory effect observable, while $Q_H^\pm$ is given by an integral of the energy momentum tensor. The horizon charge allows for the same split $Q^{\cal H} = Q^{\cal H}_S + Q^{\cal H}_H$ between a soft and a hard contribution. 

Let us now work out the hard part of the horizon charge using the equations in section 7 of~\cite{Hawking:2016sgy}. Starting from HPS (7.26) and imposing the residual horizon gauge fixing condition (7.43), one obtains for the total charge at the horizon (here we drop the superscript on $Q^{\cal H}$)
\bea
    Q_{f}\is \frac{1}{8\pi}\int_{\mathcal{H}^+}\!\!\!
    d^2z \spc dv\, f \left[\frac{1}{2M}D^A\nspc D^B\nspc \sigma_{AB} +32\pi M^2 T_{vv}^M-16\pi M D^A T_{Av}^M\right].
\eea
The first term is the soft charge $Q_S$ and the last two terms combined are the hard charge $Q_H$. The soft term generates a supertranslation on the space-time metric. In particular 
\bea
\bigl[Q_f, h_{AB}\bigr] \is 2 iM (2D_A D_B f - \gamma_{AB} D^2 f).
\eea
Hence the commutator with $Q_f$ shifts the soft hair mode $\hat{f}$ by an amount $f$. The hard part of the charge implements an active supertranslation diffeomorphism on the matter fields.

To find the explicit form of the physical operators, let us first consider the scattering states. Physical scattering states (which avoid IR divergences) must satisfy the physical state condition 
\bea
(Q_S + Q_H) \li{phys}\ra \is 0.
\eea
Projecting onto eigenstates of the respective supertranslation charges, these physical states factorize as 
\bea
\li p,z \ra_{phys} \! = \, \li p,z\ra_{\! H} \li p,z\ra_{\! S}, & &  \quad \left\{ \begin{array}{c}
{ Q_H \li p,z\ra_{\! H}  = \;  -p\spc f(z) \spc \li p,z \ra_{\! H}}\\[3mm]
\  \,{ Q_S\spc \li p,z\ra_{\! S}\;  = \spc  p f(z) \li p,z \ra_{\! S}}. \ \end{array} \right. 
\eea
Defining the corresponding matter and soft gravitational operators via \cite{Nande:2017dba}
\bea
\li p,z \ra_{\! H} =\, \mathcal{O}_{p}(z)\li 0\ra_{\! H} , \quad & & \quad \li p,z \ra_{\! S} =\, \mathcal{W}_{p}(z)\li 0\ra_{\! S}
\eea
we can associate to the state $| p,z \rangle_{phys}$ a physical operator of the factorized form 
\bea
\label{eq:wilson}
{\cal O}_{phys}(p,z) \is \mathcal{O}_{p}(z)\, \mathcal{W}_{p}(z).
\eea
Recalling that supertranslations act on the soft mode as  
$\hat{f} \to \hat{f} +f$
we read off that the gravitational Wilson line operator appearing in~\eqref{eq:wilson} takes the form~\cite{Himwich:2020rro}
\bea
\mathcal{W}(p,z)\is e^{-{ip} \hat{f}(z_k)}.
\eea
Under a large gauge transformation, the Wilson line dressing an operator of momentum $p$ acquires a phase~$e^{-ip f}$.  This is the opposite phase acquired by the light-cone momentum eigenstate $e^{ip v}$ of  the supertranslation's action on the $v$ coordinate.  

Via the above reasoning, it is not hard to see that the gravitational dressing of an infalling matter operator ${\cal O}_{in}(v,z)$ simply amounts to replacing the $v$ coordinate by the operator valued coordinate
\bea
\label{qvee}
 \hat{v} \is   v-\hat{f}(z).
\eea
This equation should be compared with the equation \eqref{patch2} for the patching function between the near horizon $(U,V)$ coordinates and the asymptotic $(u,v)$ coordinates. Concretely, if we adopt the prescription that in the near horizon region, the metric (in the local absence of infalling or outgoing matter) must be of the standard no hair form with classical coordinates $(U,V)$, then the soft hair degree of freedom $\hat{f}$ gets absorbed into the definition of the $(u,v)$ light-cone coordinates that parametrize the outside region. Since the soft mode $\hat{f}$ is now a quantum mechanical variable, this also turns the $(u,v)$ coordinates into quantum operators. From \eqref{patch2}-\eqref{patchtwo} we read off that the quantum version of the ingoing light-cone coordinate $v$ is given by \eqref{qvee}, while the quantum version of the outgoing light-cone coordinate $u$ takes the form
\bea
\hat{u} \is u\spc  - \hat{f} - \spc 
 \gamma \spc e^{(u-v)/{4M}}\spc D^2 {\hat{f}}.
\eea
The apparent asymmetry between $u$ and $v$ is due to the fact that, following HPS, we work in the advanced Bondi gauge. 

We conclude that physical operators that create early infalling modes or late outgoing modes are expressed as follows
\bea
\label{patchopv}
{\cal O}_{in}^{phys}(v,z)\! \is\! {\cal O}_{in}(\hat{v}_-,z), \qquad  \   \qquad 
{\hat{v}_-}\spc = \spc v - \spc {\hat{f}_-} \\[3mm]
{\cal O}_{out}^{phys}(u,z) \! \is\! {\cal O}_{out}(\hat{u}_+,z), \qquad    \qquad 
{\hat{u}_+} \spc = u\spc  - \hat{f}_+ - \spc 
  \gamma \spc e^{(u-v)/{4M}}\spc D^2 {\hat{f}_+}.
\label{patchopu}
\eea

\subsection{Measuring soft hair}
\vspace{-2mm}

This completes our summary of the properties of the soft hair degrees of freedom. We have argued that the soft hair is encoded in the classical transition function between the near horizon no hair coordinate region and the asymptotic Minkowski coordinate region. Secondly, we have shown that physical observables, through their gravitational dressing, acquire a direct dependence on the soft hair Goldstone mode $\hat{f}$. As seen from combining equations \eqref{patchopv} and \eqref{fpmcom}, and as we will make more explicit below, this dependence leads to exponentially growing commutators between late and early observables.

Once encoded in the classical geometry, the soft mode $f$  remains invisible to the asymptotic observer. Since $f(z)$ is a normalizable mode, its probability distribution $\rho_\psi(f)$ spreads out over its classical moduli space. Since $f(z)$ is a function that varies across the horizon, the entropy  \eqref{ssoft} stored in $f$ is proportional to the area of black hole horizon in units of the UV resolution scale~\cite{Hawking:2016msc}. A key question of interest is whether the soft hair corresponds to a hidden quantum property of a black hole or a classical property of the outside black hole space-time that will be automatically measured by an infalling outside observer. We propose that the second interpretation is the correct one.

An infalling observer who travels from the asymptotic coordinate patch to the near horizon coordinate patch will necessarily perform a measurement that projects the soft hair mode $\hat{f}$ onto an approximate eigenstate $|f\rangle_S$ with a classical eigenvalue $f$. Her observables therefore take the form 
\bea
{\cal O} \is \sum_f \, {\cal O}_f \otimes \li f\spc \ra_O\la f \spc \ri
\eea
with $
{\cal O}_f = | f\rangle_S \langle f | {\cal O}_{phys} | f \rangle_S \langle f |$
the physical observable and $|f\rangle_O$ the state of the observer associated with the classical background labeled by $f$. Without this projection on the soft hair eigen sector, the dynamical coordinates that the infalling observer uses to make localized observations acquire a macroscopic level of quantum uncertainty.

\section{Soft Hair and the Moving Firewall}
\vspace{-2mm}

The canonical structure of the soft hair degrees of freedom can be seen as the cause or manifestation of the gravitational interactions between ingoing and outgoing particles in the neighborhood of the black hole horizon. This relation can be made more explicit by considering the commutation relation between the early observable ${\cal O}_{in}^{phys}$ and the late observable ${\cal O}_{out}^{phys}$ that measure early ingoing and late outgoing particles. 

We will first consider this commutator for asymptotic observers. The near horizon interactions and commutators both grow exponentially in the time difference, indicating that the horizon is a strongly coupled region relative to the asymptotic observer. We will thus call the horizon {\it the firewall region of the asymptotic observer}. This physical notion of the firewall is different from that of AMPS. Indeed, we will see that, as opposed to the putative AMPS firewall, the observer-dependent firewall moves inwards for infalling observers.

\subsection{Soft hair, OTOCs and the firewall}
\vspace{-2mm}

It is by now an accepted truth that signal propagation through an event horizon is a non-trivial process, and that finding a microscopic description requires careful reasoning to avoid apparent contradictions between quantum principles and semi-classical expectations. The AMPS firewall argument relies on quantum information theory and applies to a maximally entangled black hole \cite{Almheiri:2012rt}. Here we recall an older firewall argument due to 't Hooft \cite{tHooft:1990fkf} and Kiem-VV \cite{Kiem:1995iy} based on gravitational shockwave dynamics. 

Gravity naturally comes into play when comparing observations in highly boosted reference frames \cite{tHooft:1990fkf}. In \cite{Kiem:1995iy} it was shown that in the context of black hole horizon, the associated gravitational shockwave interaction leads to exponentially growing commutators between late and early operators measured by an asymptotic observer. Combined with the more recent insight that this exponential growth is a manifestation of the  underlying quantum many body quantum chaos \cite{Shenker:2013pqa,Shenker:2013yza},  this Dutch version of the firewall argument is well suited for our purpose of explaining the relation between the HPS soft hair dynamics and the AMPS firewall paradox. The new perspective that we are adding is that the Lyapunov behavior is directly linked with the soft hair dynamics of the black hole.

From the commutation relation \eqref{fpmcom} between the past and future  Goldstone modes $\hat{f}_-$ and $\hat{f}_+$ and the expressions \eqref{patchopv} and \eqref{patchopu} for the light-cone coordinates $\hat{v}_-$ and $\hat{u}_+$ we deduce that the latter satisfy  non-trivial commutation relations of the form  
\bea
\label{uvcom}
\bigl[\spc \hat{v}_-(\zet) , \hat{u}_+(\zet')\spc \bigr ]\! & \simeq & \!   \gamma\, e^{(u-v)/4M} \Lambda(\zet,\zet')
\eea
where $\Lambda(\zet,\zet')$ satisfies the Green's function identity
\bea
(D^2+2)\Lambda(\zet,\zet') \is -16\pi\, \delta^{(2)}(\zet\!\spc -\!\spc \zet').
\eea
The relation \eqref{uvcom} holds at intermediate time scales of order the scrambling time. 
\begin{figure}[t]
\centering
\begin{tikzpicture}[scale=.85]
\definecolor{darkgreen}{rgb}{.0, 0.5, .1};
\draw[thick, fill, blue,thick, pattern = dots, opacity=.8,pattern = dots] (0,6) .. controls (-2.8,3.2) and (-2.8,2.8) ..  (0,0) -- (-3,3) -- (0,6);
\draw[very thick, blue] (0,0) .. controls (2.8,2.8) and (2.8,3.2) ..  (0,6);
\draw (2.1,3) node[left]{\large $O$};
\draw[thick, purple,decorate, decoration=snake,->] (.25,.25) -- (-2.75,3.25);
\draw[thick, darkgreen,decorate, decoration=snake,<-] (.25,5.75) -- (-2.75,2.75);
\draw[thick] (0,6) -- (1.5,4.5)node[right]{\; $\mathcal{I}^+$}-- (3,3 )node[right ] {$i^0$} --(1.5,1.5) node[right]{\ $\mathcal{I}^-$ }-- (0,0) -- (-3,3) -- (0,6);
\draw[thick] (-4,2) -- (-3,3) -- (-4,4); 
\draw (.22,0.15) node[right]{\mbox{\large \,${\cal O}_{in}(\hat{v}_-,z)$}};
\draw (.2,5.9) node[right]{\mbox{\large \,${\cal O}_{out}(\hat{u}_+,z')$}};
\draw [thick, decorate,decoration=snake] (0,6) -- (-4,6);
\draw [thick, decorate,decoration=snake] (0,0) -- (-4,0);
\end{tikzpicture}\vspace{-1mm}
\caption{\addtolength{\baselineskip}{-.5mm} The commutator between an operator ${\cal O}_{in}(\hat{v}_-,z)$ creating an early infalling mode and an operator ${\cal O}_{out}(\hat{u}_+,z')$ measuring a late outgoing mode exhibits Lyapunov growth due to exponentially growing collision energy  between the modes near the horizon. This commutator is generated via the coordinate shifts \eqref{patchopv} and \eqref{patchopu} and the commutator \eqref{fpmcom} between the late and early Goldstone modes $\hat{f}_+$ and $\hat{f}_-$. }
\label{fig:commutator}
\end{figure}
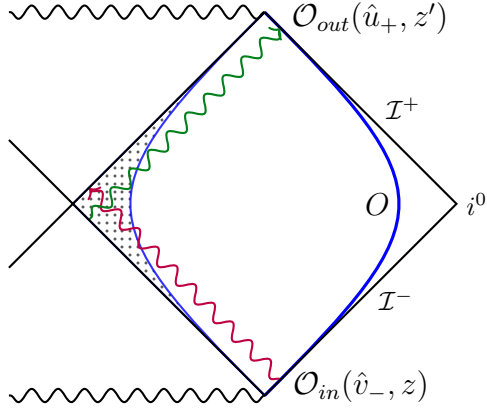
It represents a consequence of the quantum dynamics of the soft hair. The exponential growing factor matches the familiar fact that a small shift in an outgoing particle's trajectory close to the horizon can have a large effect on its future trajectory. Conversely, if we propagate an outgoing plane wave $e^{-ip_{out} \hat{u}}$ back in time, its frequency will undergo an exponential blueshift near the horizon, leading to a super-Planckian collision with the incoming modes. The physical effect of this collision is made visible via out-of-time ordered correlation functions.  

From \eqref{uvcom} and the expressions \eqref{patchopv}-\eqref{patchopu} of the physical operators, we derive that the expectation value of the (commutator)$^2$ between  early and late physical operators grows as 
\bea
\label{ophyscomm}
\frac{\Bigl\langle\, \bigl[{\cal O}_{in}(\hat{v}_-,z), {\cal O}_{out}(\hat{u}_+,z')\bigr]^2\Bigr\rangle}{\!\!
\bigl\langle \spc {\cal O}_{in}\, 
{\cal O}_{in}
\spc \bigr\rangle
\bigl\langle \spc {\cal O}_{out}\, 
{\cal O}_{out}
\spc \bigr\rangle^{}\!\!} & \! \simeq &\! \gamma^2\spc p_{in}^2\spc p_{out}^2 \spc e^{(u-v)/2M} \Lambda(z,z')^2
\eea
where $p_{in}$ and $p_{out}$ denotes the respective light-cone momenta. 
This growth continues until, after the scrambling time, the expectation value \eqref{ophyscomm} saturates at a value of order 1
\bea
\label{orderone}
\frac{\bigl\langle \bigl[{\cal O}_{in}, {\cal O}_{out}\bigr]^2\bigr\rangle}{
\bigl\langle {\cal O}_{in} {\cal O}_{in}\bigr\rangle
\bigl\langle {\cal O}_{out}{\cal O}_{out}\bigr\rangle} & \sim & 1.
\eea
In terms of the underlying ergodic microscopic theory, this growth of the (commutator)${}^2$ implies that, from the perspective of the asymptotic observer, the quantum information contained in ${\cal O}_{in}$ cannot simply pass through the horizon, but instead gets scrambled and spread out over the microscopic degrees of freedom~\cite{fastscrambler}, and soon after gets emitted in the form of Hawking radiation detectable by the outside observer \cite{HaydenPreskill}. 

The quantum dynamics mechanism that leads to equation \eqref{orderone} is that the identical operator pairs ${\cal O}_{in} {\cal O}_{in}$  and ${\cal O}_{out} {\cal O}_{out}$ in the OTOC part of the numerator cannot find a way to constructively interfere, due to the random unitary time evolution and the insertion of the other operator that separates them. This scrambling dynamics is responsible for the growing commutators, or conversely, in the gravity theory, the scrambling is an outflow of the large commutators. Indeed, from the outside perspective, the late time OTOCs are a probe of the full soft hair phase space dynamics and, due to the strongly coupled gravitational dynamics, they do not have support on some given semi-classical near horizon geometry. As we will discuss in the next subsection, this changes for OTOCs that describe the early and late observations of infalling observers who do see a given semi-classical near horizon geometry.

\subsection{Observer dependent firewall}
\vspace{-1mm}

We will now consider this same commutator between early and late operators for infalling observers following some specified observer trajectory. As we will see, the commutators in this case also grows with the time difference, and this growth again indicates the presence of a  strongly coupled region relative to the infalling observer. The effective firewall is the space-time region where the OTOCs as measured by the observer approach the saturation value \eqref{orderone}. As before, we can specify the location of the observer dependent firewall as the region where the soft hair dynamics becomes strongly coupled, relative to the probes used by the observer. As we will see, this definition of the effective firewall coincides with the one we would have guessed based on purely kinematic reasoning: it is the region within which communication with the observer necessarily involves two signals that, measured relative to each other, have super-Planckian frequencies.

For describing observations of infalling observers, it is appropriate to use the Kruskal coordinate system $(U,V)$. In the super-Planckian regime, where the longitudinal relative momentum between the incoming and outgoing waves becomes large, the waves interact via a gravitational shift interaction. In this regime we can encapsulate the interaction via a commutation relation between the two lightcone Kruskal coordinates
\bea
\label{kruscomm}
[\hat{V}_-(z),\hat{U}_+(z')] \! & \simeq & \!  \frac{\Lambda(z,z')}{F(U,V)}
\eea
where compared with \eqref{uvcom} we replaced $\gamma$ by $1/F(U,V)$, with $F(U,V)=-2g_{UV}$ for the Kruskal metric~\eqref{kruskalm}. Hence equation \eqref{kruscomm} is  covariant under redefinition of the longitudinal coordinates $(U,V)$. The Green's function $\Lambda(z,z')$ captures the transverse dependence of the shockwave interaction.

Now consider an infalling observer following some arbitrary trajectory specified via
\bea
\label{trajectory}
{\rm observer \ trajectory} \, = \, \bigl\{ \ U = X(V)\! & \leftrightarrow & \! V  = Y(U)\ \bigr\}
\eea
with $X$ and $Y$ each other's inverse: $X(Y(U)) = U$ and $Y(X(V)) = V$. Two examples of infalling observer trajectories are indicated in Figure 7. 

\begin{figure}[t]
\vspace{-1mm}
\begin{center}
\includegraphics[scale=0.65]{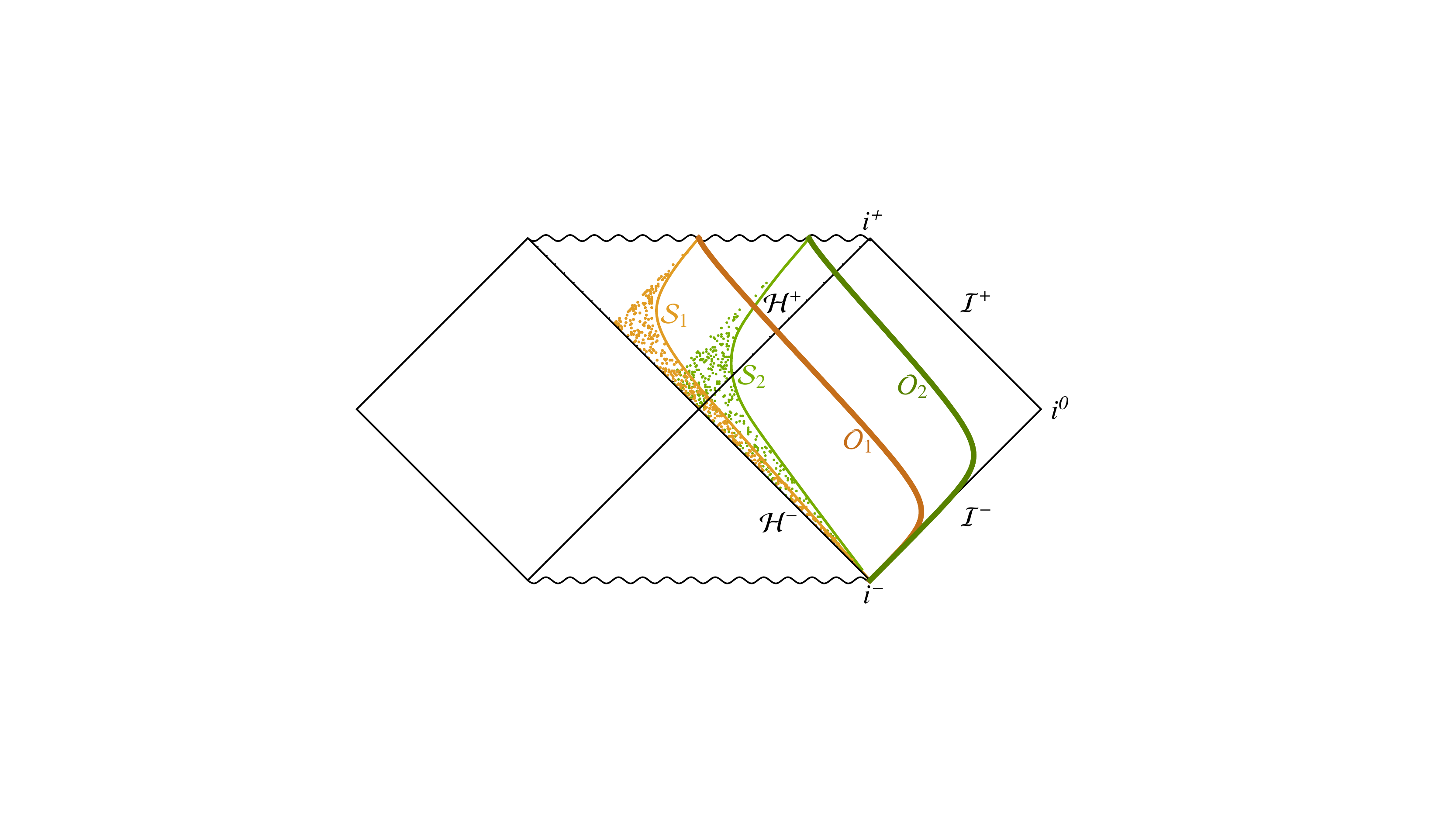}
\end{center}
\vspace{-2mm}
\caption{\addtolength{\baselineskip}{-.4mm} 
The stretched horizon ${\cal S}_i$ demarcates the region within which communication with the late observer $O_i$ requires sending super-Planckian signals. As $O_i$ falls into the black hole along the blue trajectory, the stretched horizon ${\cal S}_i$ recedes. The classical space-time region outside of ${\cal S}_i$ acts as its own physical measurement apparatus through which observer $O_i$ determines the value of the soft hair mode $f$.}
\vspace{-1mm}
\label{fig:stretchedh}
\end{figure} 

Suppose the observer sends an early signal along a light-like trajectory at constant $V$ by acting with a operator ${\cal O}_{in}$ and receives a late signal coming back along a light-like trajectory at constant $U$, measured by the operator ${\cal O}_{out}$. Assuming both operators have some approximate  frequency $\omega_{in}$ and $\omega_{out}$, as seen by the observer, the coordinate dependence of the two operators is of the form
\bea
{\cal O}_{in}(\hat{V},z)\, \sim\, e^{-i \omega_{in} \hat{\tau}_-({V},z)}, \quad & & \quad {\cal O}_{out}(\hat{U},z')\, \sim\, e^{-i\omega_{out} \hat{\tau}_+(U,z')}.
\eea
Here $\hat{\tau}_-({V},z)$ and ${\hat\tau}_+({U},z')$ are the gravitationally dressed proper time coordinates of the observer at the early and late time instances. Hence $\hat{\tau}^-$ and $\hat{\tau}^+$ are both operator valued and their mutual commutator will be responsible for the late time growth of the commutator between physical operators. Using the form \eqref{trajectory} of the observer trajectory, we have that
\bea
\frac{d\tau_-\!\!}{dV}\; =\, \sqrt{F(X(V),\nspc V)\spc X'(V)} ,\quad & & \quad
\frac{d\tau_+\!\!}{dU}\; = \, \sqrt{F(U,\nspc Y(U))\spc Y'(U)} .
\eea
Using the commutator \eqref{kruscomm} between the late and early Kruskal coordinates $\hat{U}(z')$ and $\hat{V}(z)$, we thus find that the expectation value of the (commutator)${}^2$ between the physical operators is proportional~to 
\bea
\label{newcom}
\frac{\Bigl\langle \bigl[{\cal O}_{in}(\hat V,z),{\cal O}_{out}(\hat U,z') \bigr]^2\Bigr\rangle}{\bigl\langle \spc {\cal O}_{in}\,
{\cal O}_{in}
\spc \bigr\rangle
\bigl\langle \spc {\cal O}_{out}\,
{\cal O}_{out}
\spc \bigr\rangle^{}\!\!}  &\! \simeq \! & \!  \omega_{in}^2 \omega_{out}^2 \; \Xi(U,V)\, \Lambda(z,z')^2
\eea
with
\bea
\label{xiuv}
\Xi(U,V) \! & \! \equiv\! &  \! \frac{\! F(X(V),\nspc V)\spc F(U, Y(U))}{F(U,V)^2\!}\; X'(V) \, Y'(U). \,  
\eea
Geometrically, this function $\Xi(U,V)$ equals the square of the product of the two blueshift factors between the frequency of the emitted (early) or received (late) signals and the frequency of the signals as they collide at the point $(U,V)$. The combined factor $\omega_{in}^2 \omega_{out}^2 \; \Xi(U,V)$ thus represents the square of the (center of mass energy)${}^2$ of the collision, as measured in Planck units.

The typical wave length of a physical Hawking mode, at the moment that it can be seen by an observer, is set by the size $M$ of the black hole. If we assume that both frequencies $\omega_{in}$ and $\omega_{out}$ are of order $1/M$, the collision energy becomes super-Planckian in the region where
\bea
\label{xifw}
\Xi(U,V) 
& \gtrsim &  M^4.
\eea 
This inequality defines the observer dependent firewall region bounded by the observer dependent stretched horizon. Indeed, it is easy to see that for an observer that stays outside, \eqref{xifw} coincides with the immediate neighborhood of the event horizon of the black hole, as indicated by the blue speckled region in Figure~\ref{fig:commutator}. As explained above, from the outside perspective, this firewall region is where the fast ergodic dynamics takes place, that prevents the infalling mode from truly becoming unobservable from the outside.

In Figure 7, we have indicated the observer dependent firewall region associated with two infalling trajectories, each bounded by their stretched horizon.  We see that the stretched horizon at early times extends to just outside the black hole horizon, but then recedes into the black hole interior. Outside the stretched horizon is ordinary semi-classical space-time. 
Not surprisingly, the infalling observer does not encounter her own firewall at the moment that she crosses the horizon. In geometric terms, the region inside which communication with the observer requires large blueshifts always remains at a distance, until she reaches the singularity.  In soft hair language, the infalling observer uses late time operators with only weak dependence on the late time soft mode $\hat{f}_+$, and therefore with small commutators with $\hat{f}_-$ and with the infalling mode. From the microscopic quantum perspective, her  late time operator ${\cal O}_{out}$ does not get completely scrambled relative to the early infalling operator ${\cal O}_{in}$, and vice versa. Defining such a late time operator requires undoing part of the fast scrambling time evolution. This is possible by means of a suitable approximate quantum error correction protocol and by restricting the Hilbert space to the subsector with given semi-classical near horizon geometry, i.e. to a code subspace with given value of the soft hair. The code subspace restriction is not some unusual modification of the rules of quantum mechanics, but enforced by the measurement of the classical space-time region outside of the observer dependent stretched horizon.

\def\bfU{{\bf U}}
\def\bfR{{\bf R}}

\def\smpt{\hspace{.5pt}}
\def\soft{{so\!\spc f\!\spc t}}

\section{Construction of Interior Operator}

\vspace{-2mm}

In this section we describe how the measurement of the soft hair quantum number $f$ enables the reconstruction of the Hawking partners of the late black hole radiation in the post-Page time regime. The physical mechanism is as follows: the outside measurement of $f$ projects the maximally mixed state of the black hole onto a lower entropy state contained within a code subspace ${\cal H}_f$ with given soft hair. We assume that the entropy of the code subspace $S_{code} = \log {\rm dim} \spc {\cal H}_f$ is smaller than the Bekenstein-Hawking entropy minus the entropy $S_{q\nspc f\nspc t} = \log {\rm dim} \spc {\cal H}_{q\nspc f\nspc t}$ of the interior effective QFT.
When the black hole state is no longer maximally mixed, it is possible to construct the interior modes with the help of an approximate quantum error correction protocol, that uses a recovery operator $R_f$ acting on the black hole Hilbert subspace  ${\cal H}_{\! f}$. 
The reconstruction can be performed with finite accuracy error = $e^{S_{code} + S_{q\nspc f\nspc t} - S_{B\nspc H}}\ll 1.$

\begin{figure}[t]
\begin{center}
\begin{tikzpicture}[scale=1.1]
\path[fill=green!015] (-1.3,1) -- (.025,-0.325) -- (1.5,1.15) -- (0.175,2.475)--cycle ;
\path[fill=purple!10!blue!10] (5,-.5) -- (2.43,2.07) -- (1.52,1.15) -- (4.05,-1.41)--cycle ;
\path[fill=gray!10] (-1.3,1.05) -- (-3.4,-1.05) -- (-3.16,2.98) --cycle ;
\path[fill=gray!10]  (-3.4,2.98) -- (-3.4,-1.05) -- (-3.13,2.97) --cycle ;
\draw[thick][color={rgb:green,5; black,1}]  (4.8,-.33) arc (52:120:7.25);
\draw[thick][color={rgb:green,5; black,1}]  (4.8,-.31) arc (52:120:7.25);
\draw[color={black}] [dashed] (-1.4,-1.4) -- (-3.4,.55);
\draw[color={black}] [dashed] (1.5, 2.95) -- (-2.8,-1.4);
\draw[color={black}]  (5.25,-.77) -- (1.5,2.98);
\draw[thick,decorate,decoration=snake](1.5,3.00) -- (-3.4,3.00);
\node[left] [color={blue}]  at (-.8,1.3) {\mbox{ a}};
\node[right] [color={blue}]  at (.8,1.4) {\mbox{ b}};
\node[left] [color={black}]  at (-2,1.25) {\mbox{\small Island}};
\node[above] [color={black}]  at (0,1.5) {\mbox{\large $B$}};
\node[above] [color={black}]  at (3,0) {\mbox{\large $R$}};
\node[above] [color={blue}]  at (0,-.25) {\mbox{\footnotesize Hawking}};
\node[above] [color={blue}]  at (0,-.55) {\mbox{\footnotesize pair}};
\node[right] [color={black}]  at (1.935,1.6) {\mbox{\small early}};
\node[right] [color={black}]  at (1.7,1.25) {\mbox{\small radiation}};
\draw[color={blue}][thick,->,decorate,decoration=snake] (0,0.2) -- (.95,1.15);
\draw[color={blue}][thick,->,decorate,decoration=snake] (0, 0.2) -- (-.9,1.1) ;
\draw[color={red}][thick,->,decorate,decoration=snake] (0,-1.35) -- (2.25,.9);
\draw[color={red}][thick,->,decorate,decoration=snake] (-.1, -1.3) -- (-2.1,0.7) ;
\node[color={rgb:red,5; black,1}] at (-.05,-1.45 ){\mbox{\LARGE \bf *} };
\draw[fill=black] (-1.3,1.02) circle (0.05);
\draw[fill=black] (1.5,1.13) circle (0.05); 
\end{tikzpicture}
\vspace{-3mm}
\end{center}
\caption{\addtolength{\baselineskip}{-.5mm} A late time slice through an evaporating black hole space-time. The blue region represents the early Hawking radiation $R$ and the green region is described by some dual strongly coupled CFT, labeled $B$.  For a black hole past its Page time, $B$ is maximally entangled with $R$. Modes inside the gray `Island region' have in their past collided with outgoing modes inside $R$. A late Hawking pair $a$ and $b$ is contained inside the entanglement wedge of CFT. The interior mode $a$ is decoupled from the early radiation $R$; its reconstruction via CFT operators involves a projection on a code subspace with given soft hair.  }
\vspace{0mm}
\end{figure}
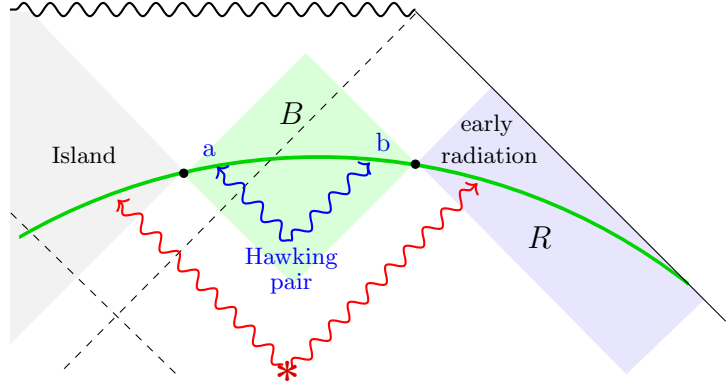

\subsection{Initial state with entangled soft hair}
  \vspace{-2mm}

Let ${\cal H}_B$ denote the total black hole Hilbert space at some given time $t$. As we have argued, the total black hole Hilbert space decomposes into a direct sum of sectors ${\cal H}_f$ with given soft hair quantum numbers
\bea
{\cal H}_B \is \bigoplus_f \; {\cal H}_f.
\eea
We will call the Hilbert spaces ${\cal H}_f$ the code subspace labeled by the soft hair eigenvalue $f$.
For simplicity, we assume the eigen sectors ${\cal H}_f$ are all isomorphic to some given abstract code Hilbert space ${\cal H}_{C}$ spanned by a set of basis states $|\spc i \rangle_C$, where $i = 1, ..., d_C$ with $d_C = {\rm dim} \spc {\cal H}_{C}$  the dimension of the code subspace.

We now also introduce an abstract Hilbert space ${\cal H}_S$ spanned by the soft hair eigenstates $|f\rangle_S$ with $f = 1,..., d_S$ where $d_S = {\rm dim} \spc {\cal H}_S$. We can then define an embedding tensor $
\tT \ : \ {\cal H}_{C} \otimes {\cal H}_{S} \ \hookrightarrow \ {\cal H}_B.$
On the basis vectors $|f\rangle_S$ with given soft hair quantum number, this tensor map defines an injective embedding ${\tT}_{\!\!\spc f}$  of the abstract code space ${\cal H}_C$ into a physical code subspace ${\cal H}_f$ inside the black hole Hilbert space
\bea
\label{tdef}
{\tT}\; \li \spc \phi \spc \ra_C  \spc \li f\ra_S\! \is\! {\tT}_{\!\!\spc f} \spc \li \spc \phi\spc \ra_C \, \in\, {\cal H}_f \, \subset \, {\cal H}_B.  
\eea
For later use, we also introduce the Hermitian conjugate   $\tT_{\!\!\spc f}^{\dag}$ of the tensor $\tT_{\!\!\spc f}$ as the inverse mapping from ${\cal H}_f$ into ${\cal H}_C \otimes {\cal H}_S$, defined via the relation
\bea
\label{tdag}
\tT_{\!\! \spc f}^{\dag} \tT_{\! f} \li \phi\ra_C \is \li \phi\ra_C \li f\ra_S.
\eea
The linear map ${\tT}^\dag_{\!\!\spc f}$ vanishes on the orthoplement of ${\cal H}_f$. In other words, $\Pi_f = \tT_{\!\! \spc f} \tT_{\!\! \spc f}^{\dag}$ is a projection operator onto the code subspace ${\cal H}_f$.

To describe the Hilbert state of a post-Page time black hole with soft hair, we start from a maximally entangled state of $ {\cal H}_{C} \otimes {\cal H}_S$ with the Hilbert space ${\cal H}_R$ of early radiation and the Hilbert space ${\cal H}_O$ of an external observer. By construction, we define the soft hair Hilbert space ${\cal H}_S$ and the Hilbert space ${\cal H}_O$ of the observer such that each forms the purification of the other.  
In other words, we only consider the soft hair quantum numbers in our discussion that have been measured by the outside observer. All other quantum degrees of freedom of the black hole are included in the code Hilbert space and assumed to be entangled with the early radiation. Using the diagrammatic notation of \cite{Yoshida:2019kyp}, we write
\bea
\li \Psi_0\ra\! \is \! \frac{1}{\sqrt{d_C\spc d_S}} \, \sum_{i,f} \; \li \spc i\spc  \ra_C\spc \li f\ra_S \spc \li f\ra_O\spc \li \spc i \spc\ra_R\ = \ \begin{tikzpicture}[xscale=0.65,yscale=.65,baseline=(current bounding box.center)]
    \draw[thick,->] (0,0)--(1.75,0)--(1.75,0.4)node[above right] {\scriptsize $\!R$}--(1.75,1.4);
      \draw[thick,->] (0,0)--(-1.75,0)--(-1.75,0.4)node[above left] {\scriptsize $C\!$}--(-1.75,1.4);
      \filldraw [](0,0) circle (2pt);
    \draw[thick,->] (0,.4)--(.65,0.4)--(.65,0.5)node[above right] {\scriptsize $\!O$}--(.65,1.4);
      \draw[thick,->] (0,.4)--(-.65,.4)--(-.65,0.5)node[above left] {\scriptsize $S\!$}--(-.65,1.4);
      \filldraw [](0,0.4) circle (2pt);
    \end{tikzpicture}.
\eea

Next, we use the tensor $\tT$ to map this auxiliary entangled state
into a maximally entangled state of the black hole with the early radiation and the observer. The result will be our initial state $|\Psi(0)\rangle$ to which we will then apply the Hawking time evolution and holographic reconstruction. 
In diagrammatic notation 
\bea
 \li \Psi(0)\ra\!\! \is\!\! \tT\spc \li \Psi_0\ra 
 \ \, =\ \; \raisebox{2mm}{$
\begin{tikzpicture}[xscale=0.65,yscale=.55,baseline=(current bounding box.center)]
    \draw[thick,->] (0,0)--(1.75,0)--(1.75,1.8)node[above right] {\scriptsize $\!R$}--(1.75,2.8);
      \draw[thick] (0,0)--(-1.75,0)--(-1.75,0.45)node[above left] {\scriptsize $C\!$}--(-1.75,1.4);
      \filldraw [](0,0) circle (2pt);
    \draw[thick,->] (0,.4)--(.65,0.4)--(.65,1.8)node[above right] {\scriptsize $\!O$}--(.65,2.8);
      \draw[thick] (0,.4)--(-.65,.4)--(-.65,0.45)node[above left] {\scriptsize $S\!$}--(-.65,1.4);
      \draw[thick,->] (-1.2,2)--(-1.2,2.7)node[right] {\scriptsize \,$B$}--(-1.2,2.8);
      \draw[thick,fill=green!0] (-2.2,1.2) -- (-2.2,2.2) -- (-0.2,2.2) -- (-0.2,1.2) --cycle  ;
\draw (-1.2,1.7) node {\small ${\tT}$};
      \filldraw [black](0,0.4) circle (2pt);
    \end{tikzpicture}$}
\eea
or in an equation
\bea
\label{psiexp}
\li \Psi(0)\ra \is \frac{1}{\sqrt{d_C\spc d_S}} \, \sum_{i,f}\; {\tT}_{\!\! \spc f} \spc \li \spc i\spc \ra_C 
\otimes \spc  |f\ra_{\! O}\spc \li\spc i\spc \ra_{\! R} .
\eea
This is our initial state of the post-Page time black hole with its soft hair entangled with the observer $O$.

\vspace{-1.5mm}

\subsection{Time evolved state with entangled soft hair}
  \vspace{-1mm}

Next we apply the time evolution $\uU_{\! \tau}$ over some finite time  interval $\tau$ of order the black hole scrambling time $\tau_s \simeq M \log M$. During this time, the black hole emits some additional Hawking radiation. Let ${\cal H}_b$ denote the Hilbert space of the late radiation, and let
${\uU} = {\uU}_{\! \tau}\spc \circ {\tT}$
denote the combined operation of embedding ${\cal H}_{C} \otimes{\cal H}_S$ into ${\cal H}_{B}$ and then applying unitary time evolution 
\bea
\label{psievol}
\li \Psi(\tau)\ra \is \uU \spc \li \Psi_0\ra\, = \,  {\uU}_{\! \tau} \li \Psi(0)\ra . 
\eea
We can diagrammatically represent the entangled state of the time evolved black hole as in the figure below

\bigskip

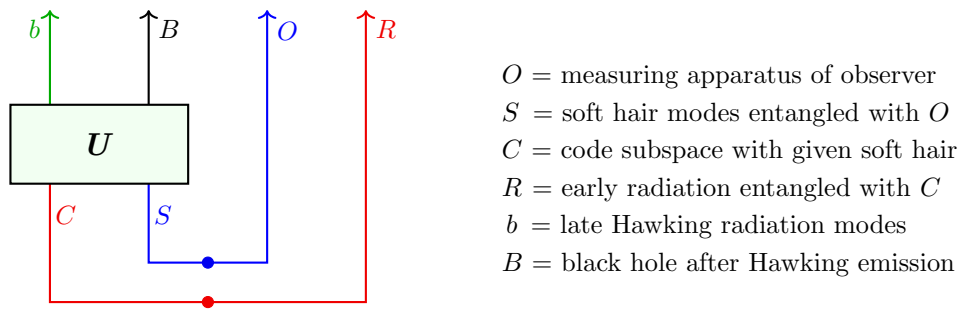
\begin{figure}[hbtp]
\begin{center}
\begin{tikzpicture}[xscale=1.05,yscale=1.05]
\draw[thick,->][color={rgb:red,3; black,.2}](1,3)--(1,1.6)node[right] {$\!{C}$}--(1,0.5)--(5,0.5)--(5,4.2) node[below right] {$R$};
\draw[thick,->][color=blue] (2.25,3)--(2.25,1.6)node[right] {\!$S$}--(2.25,1)--(3.75,1)--(3.75,3.93) node[right] {$O$}--(3.75,4.2);
\draw[thick,->][color={rgb:green,2; black,1}](1,2.35) -- (1, 4.2) node[below left] {$b$};
\draw[thick,->](2.25,2.5) -- (2.25, 4.2)node[below right] {$B$};
\draw[thick] [fill=green!5](.5,2) -- (2.75,2) -- (2.75,3) -- (.5,3)  --cycle ;
\draw (1.625,2.5) node {\large ${\uU}$};
\filldraw [color={rgb:red,3; black,.2}](3,.5) circle (2pt);
\filldraw [color={rgb:blue,3; black,.2}](3,1) circle (2pt);
\end{tikzpicture}
~~~~~~~~~\begin{tikzpicture}
\draw[thick, bag,align=left,font=\linespread{0.9}\selectfont] (5+1,.25+2.5) rectangle (5+1,.25+2.5) node []
{$O$ = measuring apparatus of observer\\[1.2mm] 
$S\,$ = soft hair modes entangled with $O$\\[1.2mm] 
$C$ = code subspace with given soft hair \\[1.2mm] 
$R\spc$ = early radiation entangled with $C$\\[1.2mm] 
${}$\,$b\,$~= late Hawking radiation modes\\[1.2mm] 
$B$ = black hole after Hawking emission\\};
\end{tikzpicture}
\end{center}
\vspace{-1mm}
\caption{Maximally entangled state of a post-Page time black hole after a period of Hawking evaporation. If $\uU$ is scrambling and $d_S \gg d_b$, the state of the late and early radiation looks like a factorized product state $\rho_b \otimes \rho_R$. The reconstruction of the black hole interior must involve the Hilbert space of the observer.
}
\vspace{0mm}
\end{figure}

\noindent
The situation depicted in Figure 8 needs to be compared with that of the standard firewall discussion. The new ingredient relative to AMPS is that we have divided the black hole Hilbert space into different soft hair sectors that are entangled with the outside observer $O$. 

Plugging \eqref{psiexp} into \eqref{psievol} and expanding into basis states $|n\rangle_b$ of the late radiation, we can write $|\Psi(\tau)\rangle$~as
\bea
\li \Psi(\tau)\ra 
\is \sum_{i,f,n} \,\frac{1}{\sqrt{d_C\spc d_S}\!}\;  {\bfC}_{\!\spc n} {\tT}_{\!\! f}\spc \li\spc i\spc \ra_{\! C}\otimes \li n\ra_{\nspc b} \spc  \li\spc i\spc \ra_{\! R}\spc \li f\ra_{\! O} .
\eea
The $\bfC_n$ are called Kraus operators. In the following, they are assumed to be random ergodic operators, subject only to the macroscopic conservation laws and the unitarity relation
\bea
\label{cunitarity}
\sum_n {\bfC}_{n}{\!\!\!\spc}^\dag \, {\bfC}_{\! \spc n} \is {\mathbb 1}_{{}{B}}.
\eea

The idea of including the Hilbert space of an infalling observer to resolve the firewall obstruction is not new. Our discussion below closely parallels that in recent work \cite{Yoshida:2019kyp} of Yoshida, with the main physical difference that \cite{Yoshida:2019kyp}  incorporates the Hilbert space of the observer as part of the interior black hole dynamics, while in our set up the observer Hilbert space is a separate sector outside the black hole. Our soft hair Hilbert space plays the same formal role as the observer Hilbert space of \cite{Yoshida:2019kyp}.

The introduction of the observer Hilbert space has important consequences for the reconstruction of the Hawking partners of $b$. In the AMPS set up, the late radiation $b$ is argued to be completely entangled with the early radiation $R$. This is no longer the case in our situation. In \cite{Yoshida:2019kyp}, Yoshida proves the following 

\noindent
{\bf Decoupling theorem (Yoshida):} {\it If the unitary operator $\uU$ is scrambling and the dimension of the soft hair Hilbert space is much larger than the dimension of the late radiation Hilbert space: $d_S \gg d_b$, then the late radiation $b$ and early radiation $R$ are decoupled (not entangled). The combined density matrix of late and early radiation factorizes into an approximate tensor product $
\rho_{Rb}  \simeq\, \rho_{R} \otimes \rho_b$
with error of order $d_b^2/d_S^2$.}

\noindent
The immediate implication of this decoupling theorem is that, unlike in the AMPS scenario, it is not possible to purify the late Hawking radiation by means of the early radiation. Indeed, theorem 1 in \cite{Yoshida:2019kyp} continues: 

\noindent 
{\it Furthermore, for any operator ${\cal O}_b$ on the late radiation b, a
partner operator $\widetilde{\cal O}_b$ can be constructed on ${\cal H}_B$ and ${\cal H}_O$ without using any degrees of freedom from the early radiation~R.}

This conclusion is in accord with normal physical expectations. As we will make explicit below, the only role of the Hilbert space of the observer is to first identify soft hair sector of the black hole. This happens automatically via the gravitational dressing of the effective QFT observables, in combination with the fact that the observer herself is also entangled with the same soft hair degrees of freedom. This is a standard, though usually implicit, step in constructing quantum observables: any standard quantum observable only makes physical sense within some given classical environment.

\vspace{-1mm}

\subsection{Density matrix of black hole and observer}

\vspace{-1mm}

Before describing the recovery map and interior operators, let us write the explicit form of the density matrix of the black hole and the observer, obtained by tracing out the early and late radiation. It decomposes as a sum over soft hair sectors 
\bea
\rho_{B\spc O}(\tau) \is  \tr_{{}_{\mbox{\scriptsize ${\cal H}_R\! \otimes\! {\cal H}_b$}}}\! \Bigl(\spc |\Psi(\tau)\rangle \langle \Psi(\tau) |\spc \Bigr) \, = \, \sum_f\; \sigma_{B\nspc f} \spc \otimes\spc | f\rangle_O\langle f | .
\eea
Here $\sigma_{B\nspc f}$  denotes the density matrix of the black hole with given soft hair quantum number $f$. It can be expressed in terms of the Kraus operators as
 \ba
 \label{sigmarad}
 \sigma_{B\nspc f}\!\is \! \frac{1}{d_C} \, \sum_n  \bfC_{\nspc n}  \Pi_f \spc  
 \bfC^\dag_n
 \ea
with $\Pi_f = \tT_{\!\!\spc f}\spc  \tT^\dag_{\!\!\spc f}$ the projection onto the code subspace ${\cal H}_f$. The density matrix $\sigma_{B\nspc f}$ acts as a projection operator onto the time evolved embedding of code subspace ${\cal H}_f$ with fixed soft hair~$f$. Indeed, an equivalent characterization of $\sigma_{B\nspc f}$ is as the density matrix obtained through the application of a quantum noise channel to the normalized unit density matrix on the abstract code subspace ${\cal H}_C$
 \ba
 \label{noisechannel}
\quad \sigma_{B\nspc f} = \, {\cal N}_f(\sigma_{C})\! \is \!  \tr_{\spc{\cal H}_b}\Bigl( \uU \bigr(\spc \sigma_{C}\otimes | f\rangle_{S}
 \langle f |\spc \bigr) \uU^\dag\Bigr), \qquad \ \ \sigma_C \equiv \, \frac{1}{d_C} \,\mathbb{1}_C .
\ea 
Here ${\cal N}_f$ denotes the noise channel specified by the Hawking evolution process $\uU = \uU_\tau \circ \tT$ applied to the initial mixed state in ${\cal H}_{C} \otimes {\cal H}_S$ with given initial soft hair quantum number $f$.

\vspace{-1mm}
\subsection{The recovery map}

  \vspace{-1mm}
  
We now describe the recovery map that enables the construction of the Hawking partners of $b$ by means of an operator acting only on ${\cal H}_B$ and ${\cal H}_O$. The procedure is a slight improvement of the one presented in \cite{Verlinde:2012cy} and makes use of an approximate quantum error correction protocol similar to that used in the Petz map \cite{Cotler:2017erl, work-in-progress}.

The recovery map $\rR$ attempts to reverse the unitary operator $\uU$ in \eqref{psievol}. However, it can not do so exactly, because it does not have access to the $b$ Hilbert space. To overcome this obstacle we introduce ancillary mirror degrees of freedom $a$, that temporarily stand in for the physical Hawking partners of the late radiation $b$. The recovery map $\rR$ acts on the black hole and the ancillary Hilbert space ${\cal H}_a$ and simultaneously accomplishes three interrelated tasks: it
i) recovers the quantum information contained in the code Hilbert space, ii) performs an entanglement swap between the Hawking partners of $b$ to the ancillary space ${\cal H}_a$, iii) enables the construction of interior operators $O_a = \widetilde{O}_b$ that act on the Hawking partners of $b$.

\begin{figure}[t]
\begin{center}
\begin{tikzpicture}[xscale=1,yscale=1]
\draw[thick](3,3.8) -- (3.65,3.8)  
-- (4, 3.8);
\draw[thick,->][color={rgb:red,3; black,.2}](1,2)--(1,1.3)node[right] {${C}$}--(1,0.5)--(5.5,0.5)--(5.5,4.9) node[below right]{$R$};
\draw[thick,->][color=blue] (2.4,2)--(2.4,1.3)node[right] {$S$}--(2.4,1)--(4,1)--(4,3.05) node[right] {\!$O$} -- (4,4.7) node[right] {$f$}--(4,4.9);;
\draw[thick,->][color={rgb:green,2; black,1}](1,2.35) -- (1, 4.9) node[below left] {$b$};
\draw[thick,->][color={rgb:cyan,2; black,1}](1.8,4) -- (1.8, 4.85) node[below right] {$a$}-- (1.8, 4.9);
\draw[thick](2.375,2.5) -- (2.375, 3.3)node[below right] {\!$B$} -- (2.375, 3.4);
\draw[thick,->][color={rgb:red,3; black,.2}](2.85,3.5) -- (2.85, 4.9)node[below right]  {${C}$};
\draw[color={rgb:gray,0; black,5}] (1.42,5.1) node {\scriptsize $|\rm TFD\rangle$};
\draw[thick][fill=green!5] (.5,1.75) -- (3,1.75) -- (3,2.6) -- (.5,2.6) --cycle  ;
\draw[thick][fill=red!5] (1.5,3.4) -- (3.3,3.4) -- (3.3,4.2) -- (1.5,4.2) --cycle  ;
\draw (1.65,2.2) node {\Large $U$};
\draw (2.35,3.8) node {\Large $R_f$};
\draw[thick] [fill=red!5] (4,3.8) circle (2.5pt);
\draw[thick] [fill=red!5](3.8,3.4) -- (4.2,3.4) -- (4,3.7) -- (3.8,3.4) --cycle ;
\filldraw [color={rgb:red,3; black,.2}](3.25,.5) circle (2pt);
\filldraw [color={blue}](3.25,1) circle (2pt);
\end{tikzpicture}
\end{center}
\vspace{-1mm}
\caption{The recovery operator $\bfR$ acts on the tensor product of the black hole Hilbert space $B$ and the observer Hilbert space $O$. The final state after the recovery operation consists of the tensor product of the initial state $|\Psi\rangle_{C}$ and the thermofield double state of the late radiation $b$ and ancilla $a$.}
\end{figure}
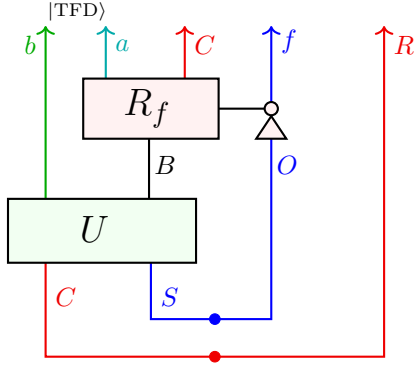

The recovery map $\rR$ is designed to approximately recover the initial state and in the process put the late radiation $b$ and ancillary degrees of freedom $a$ into a thermofield double state. As we will show below, the sequential action of the time evolution $\uU$ and the recovery operation $\rR$ yields\footnote{Here implicitly extended  $\rR$ and $\uU$ to operators $\rR \otimes \mathbb{1}_{{b}}$ and $ \uU  \otimes \mathbb{1}_{a}$ acting on 
  ${\cal H}_{B} \! \otimes \!{\cal H}_{b}\!\otimes \!{\cal H}_{a}$.}
  \ba
\label{RecTFD}
\rR \spc \uU \, |\Psi_0\rangle \! &  \simeq & \! |\Psi_0\rangle \ |{\rm TFD}\rangle_{ab} . 
 \ea
The $\,\simeq\,$ sign indicates that the recovery has a small error of order 
\bea
\label{accuracy}
\frac{d_{C}d_{b}}{d_{{B}}}\!  \is \! e^{S_{{\rm code}} + S_{{b}} - S_{B}}\ll~1.
\eea

The key observation that enables the recovery operation is that the matrix elements of the Kraus operators $\bfC_n$ can be treated as random matrices. This property implies that the $\bfC_n$ look invertible when restricted to act within a code subspace. In a holographic setting like ours, the code subspace is defined to contain the Hilbert space of the low energy QFT in some given semi-classical background geometry. In our case, the classical geometry and the corresponding code subspace are both labeled by the soft hair quantum number~$f$. 

To define a state independent operator $\rR$, we allow it to act both on the black hole Hilbert space and the Hilbert space of the observer. The state of the observer is assumed to be maximally entangled soft hair degrees of freedom. So ${\cal H}_O$ is spanned by states $|f\rangle_O$, that are uniquely paired with the black hole state with a given soft hair quantum number $f$. Correspondingly, we expand $\rR$ as a sum of recovery operators $\rR_f$ acting within a fixed soft hair sector via
\bea
\rR \is \sum_f\, \rR_{\nspc f}\,
\otimes \li f\ra_{O}
 \la f\ri .
\eea
As we did for the time evolution operator $\uU$, we decompose the recovery map $\rR_f$ in terms of Kraus operators
\ba
\label{recovery}
{\rR}_f \spc\li \Phi \ra_{{\!}{B}}
 \is \sum_n\,  {\rR}_{\nspc f,n} \spc \li \Phi \ra_{{\!}{B}}\otimes \li f \ra_S\spc 
 \li \smpc n \smpc\ra_{a} .\ea
 The operators $\rR_{f,n}$ are designed to reverse the action of the Kraus operators $\bfC_n$. 
Inspired by the form of the Petz map \cite{Cotler:2017erl}, we choose the individual recovery operators as follows\cite{work-in-progress}
\ba
\label{recndef}
 {\rR}_{\nspc f,n} \is \frac{1}{\sqrt{d_C}}\tT_{\!\!\spc f}^{\dag} \spc {\bfC}^{\spc\dag}_{\nspc n}\, \sigma_{B\!\spc f}^{-1/2} 
\ea 
with  $T_f^\dag$ and $\sigma_{B\!\spc f}$ given in \eqref{tdag} and \eqref{sigmarad}.
By construction, this operator satisfies the following exact~identity
\bea
\label{rchannel}
{\cal R}_f(\sigma_{B\nspc f})\,  \equiv\,  \tr_{{\cal H}_a}\! \Bigl( \rR_{\nspc f}\spc \sigma_{B\nspc f} \spc \rR_{\nspc f}^\dag\Bigr)\! \is\! \sigma_{C}\otimes \li f\ra_{S}
 \la f \ri .\spc
\eea
In words, the reverse noise channel ${\cal R}_f$, defined by applying the recovery operator $\rR_f$ and then tracing over the ancillary Hilbert space ${\cal H}_a$, applied to the black hole density matrix $\sigma_{B\nspc f}$ at time $t=\tau$ reproduces the initial density matrix $\sigma_C$ on the code subspace. Equation \eqref{rchannel} directly follows from the definition \eqref{recndef} and the unitarity relation for the Kraus operators.
For our purpose, the important property of the recovery operator $\rR$ is that when applied to any final state $|\Psi(\tau)\ra$ it satisfies the approximate identity \eqref{RecTFD}.

{The proof of \eqref{RecTFD} exploits the fact that the Kraus operators are random matrices subject to the unitarity relation 
\eqref{cunitarity}. 
Basic statistical reasoning then shows that they satisfy the following  approximate identities
\bea
\label{statrel}
{\bfC}^\dag{\!\!\!}_m\spc \sigma_{B\nspc f}^k \spc {\bfC}_{\!\smpc n}  \! & \simeq & \! \frac{1}{d_C^k}\, {p^{k+1}_n} \spc \delta_{mn}\; {\mathbb 1}_{{}{B}}
\eea
with $p_n$ the Boltzmann weights.
This set of identities hold with finite accuracy \eqref{accuracy} and follow straightforwardly from the explicit form \eqref{sigmarad} of $\sigma_{B\nspc f}$ and by repeated application of the special $k=0$ identity 
\bea
{\bfC}^\dag{\!\!\!}_m\spc  {\bfC}_{\!\smpc n}\!&\!  \simeq \!&\!  p_n\spc \delta_{mn}\; {\mathbb 1}_{{}{B}}. 
\eea
Using the special case of \eqref{statrel} for $k=-1/2$ gives
\bea
\rR_{f,n} \bfC_m \tT_{\! f} \li \phi\ra_C \! & \simeq  & \! \sqrt{p_n\!}\; \delta_{nm}\, \li \phi\ra_C \li f\ra_S .
\eea
From this, we easily verify that
\bea
\rR \spc \uU \, \li\Psi_0\ra 
\is \frac{1}{\sqrt{d_C d_S}}\, \sum_{n,m,i,f} \rR_{f,n} \bfC_m \tT_{\! f}\, \li\spc i \spc \ra_C\otimes\li f\ra_O  \li \spc i \ra_R \li n\ra_a\li m \ra_b \nonumber\\[-2mm]\\[-2mm]\nonumber
& \simeq &\! \frac{1}{\sqrt{d_C d_S}}\, \sum_{i,f}\; \li\spc i \spc \ra_C \li f\ra_S \li f\ra_O\li \spc i \ra_R  \otimes \sum_n \sqrt{p_n\!}\; \li n\ra_a\li n \ra_b\, \simeq \, \li \Psi_0\ra \otimes \li {\rm TFD}\ra_{ab} .
\eea}
\def\ttwo{{\mbox{\scriptsize $2$}}}

\vspace{-2mm}

\subsection{Interior operators}
\vspace{-2mm}

We now briefly review the construction of the interior operators $\widetilde{O}_{a}$ that create or annihilate the Hawking partners of the late radiation modes $b$. These operators act on the the tensor product of the black hole Hilbert space ${\cal H}_B$ and the observer Hilbert space ${\cal H}_O$. Geometrically, they live inside the interior region just behind the black hole horizon still within the CFT region of the black hole interior, as indicated in Figure~\eqref{fig:stretchedh}. The key idea is to use the property 
\eqref{RecTFD} of the recovery map to define operators $\widetilde{O}_{a}$ such that the time evolved state $|\Psi(\tau)\rangle$ looks like the local TFD vacuum state $|{\rm TFD}\rangle_{ab}$ in terms of the interior $a$ and exterior $b$ modes. The following definition of the mirror operator is a small modification of the reconstruction based on approximate error correction developed in \cite{Verlinde:2012cy} during the early days of the firewall debate.

The recovery operator $\rR$ can be used to associate to any operator ${\cal O}_a$ acting on the ancillary Hilbert space ${\cal H}_a$ a mirror operator acting on ${\cal H}_B\otimes {\cal H}_O$ via 
\bea 
\widetilde{\cal O}_{a}\, = \, \rR^\dagger \, {\cal O}_a\, \rR \is \sum_f\; \widetilde{\cal O}_{a,f} \otimes
\li f\spc \ra_O \la f\spc\ri\\[1mm]
\widetilde{\cal O}_{a,f} \, = \,  \rR_f^\dagger \, {\cal O}_a\, \rR_f \! \is \sum_{n,m} \;  \raisebox{-2pt}{${}_a$}\!\la m \ri {\cal O}_a\li n\ra_{\! a}\, \rR^\dag_{f,m} \rR_{f,n} .\,
\eea
The properties of the mirror operators are analyzed in some more detail in \cite{Verlinde:2012cy}. In particular, using the result \eqref{RecTFD}, we immediately read off that the expectations values of interior and exterior QFT operators reproduce those of the local Unruh vacuum state 
\bea
\la \Psi(\tau) \ri\, {\widetilde{\cal O}}_{a}^{\spc (1)}\,  {\cal O}^{(2)}_b \, \li \Psi(\tau)\ra \is {}_{\raisebox{-2pt}{\scriptsize $ab$}}\!\la {\rm TFD}|\, {\cal O}_a^{(1)} \, {\cal O}_b^{(2)}\, \li {\rm TFD} \ra_{\! ab} .
\eea
Moreover, using the random matrix properties of the Kraus operators $\bfC_n$, it can be shown that the mirror operators ${\widetilde{\cal O}}_{a}$ satisfy the identical operator algebra as the corresponding low energy field theory operators ${{\cal O}}_{a}$, up to corrections of order \eqref{accuracy}. Hence as long as the code space remains small enough, this resolves the firewall paradox.

The above reconstruction of interior operators relies on the fact that the black hole evaporation process is described by a random unitary dynamics on the full black hole Hilbert space, including the soft sector. So even if the system starts in a small enough code subspace, over time the system will evolve into a mixed state with maximal entropy saturating the BH bound. The interior effective QFT then breaks down. As anticipated in \cite{Verlinde:2012cy}, this breakdown is a consequence of an overzealous attempt to capture a close-to-maximally entangled state of the total black hole system in terms of a single semi-classical reality. Instead, the maximally mixed state represents an incoherent sum of different semi-classical states that can be distinguished by an appropriate outside measurement.

\medskip

\section{Conclusion}
\vspace{-1mm}

We have studied the soft hair degrees of freedom associated to the global black hole space-time and connected it to the discussion of the AMPS firewall paradox. Following HPS, we used the structure of the asymptotic BMS symmetry algebra as a guide for extracting the dynamics and phase space structure of soft hair modes at the horizon. 

We then saw that the soft hair phase space plays multiple roles.  It labels the classical background on which an observer can make measurements, dynamically influences the external geometry by capturing shifts in the horizon due to the back-reaction of infalling matter, and contributes to the entropy of the black hole. While it has been proposed that tracing over these soft modes (which include superrotations~\cite{Haco:2018ske} and gauge transformations~\cite{Hawking:2016msc}) may be enough to produce the full black hole entropy, for our purposes, we only need its dimension to be large enough to include the effective QFT degrees of freedom.   A key part of our story is that the infalling observer, rather than tracing over soft modes, performs a measurement of the soft degrees of freedom.

The fact that the soft hair modes are a classical manifestation of the black hole entropy and that their dynamics exhibits Lyapunov behavior  are related. There is a close connection between Lyapunov behavior and decoherence \cite{Zurek:1994wd,Zurek:1995jd}. In quantum systems that exhibit classical chaos, the von Neumann entropy tends to grow linearly in time at a rate proportional to the Lyapunov exponent. This growth is caused by the breakdown of Liouville's theorem in the quantum setting: the phase space probability distribution expands in one direction, while the shrinking width in the dual direction quickly reaches an equilibrium due to quantum diffusion. The linear entropy growth indicates that the system decoheres: the interaction with the environment produces an incoherent superposition of approximate eigenstates of the Lyapunov mode. In our case, this growing mode is the soft hair Goldstone field $f(z)$.

 Soft hair modes are part of the semi-classical  space-time experienced by an infalling observer. While an observer has the freedom to gauge fix her coordinate system,  by falling into the black hole her observations are sensitive to the non-trivial gauge-invariant observables that arise in the presence of a black hole horizon. Our proposal is that the semi-classical background space-time experienced by an observer acts as a measurement apparatus for the soft modes. We used the dynamical properties of the soft hair to identify the semi-classical location of the observer dependent firewall and found that the infalling observer does not encounter this firewall before reaching the singularity. Rather, the firewall recedes into the interior as the observer falls in, revealing a classical space-time accessible to low energy observations.
 
Finally, we argued that the soft hair degrees of freedom form a quantum resource to disentangle early and late Hawking radiation and thereby avoid the firewall paradox. The soft hair's entanglement with the state of an infalling observer is what allows us to define operators that reconstruct the Hawking partners associated to late Hawking radiation and preserve the entanglement structure necessary for the infalling observer to experience a smooth horizon.

\vfill

\subsection*{Acknowledgements}
\vspace{-1mm}

We thank Ahmed Almheiri, Laura Donnay, Steve Giddings, Daniel Harlow, Yasunori Nomura, Geoff Penington,  Andrea Puhm, and Erik Verlinde for valuable discussions and comments.  The research of SP is supported by a fellowship at the Princeton Center for Theoretical Science.  The research of HV is supported by NSF grant number PHY-1914860.  

\bigskip

\pagebreak

\appendix

\section{Soft hair phase space redux}\label{softphasespace}
\vspace{-2mm}

In this Appendix we will examine the phase space of the boundary modes associated with the horizon and null infinity in more detail. Our specific aim is to show that the presence of the black hole horizon indeed leads to additional soft hair degrees of freedom relative to asymptotically flat Minkowski space. Since the existence of black hole soft hair (or more accurately, the relevance of asymptotic BMS symmetries for the quantum properties of black holes) was in fact disputed in follow up work to HPS \cite{Mirbabayi:2016axw, Bousso:2017dny}, it is worth spelling out how they arise in our geometric set up.

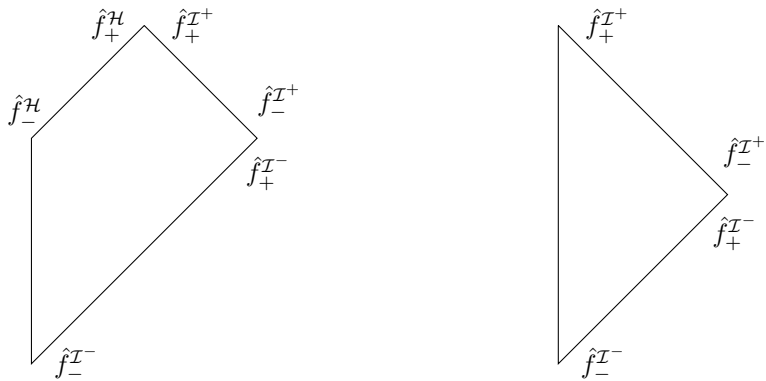
\begin{figure}[b]
\vspace{-4mm}
\begin{tikzpicture}[scale=.5]
\definecolor{darkgreen}{rgb}{.0, 0.5, .1};
\draw (0,6) 
-- (2.75,3.25 )node[above right ]{\!\! $\hat f_-^{{\cal I}^+}$} -- (3,3) -- (2.8,2.8) node[below right ]{\!\!\!\!\! $\hat f_+^{{\cal I}^-}$} --
(0,0)--
(-3,-3)node[right]{\, $\hat f^{\mathcal{I}^-}_-$} -- (-3,3)node[above]{$\hat f^{\cal H}_{-}\ \,$}-- (0,6)node[left]{$\hat f^{\cal H}_{+}\ $}-- (0,6)node[right]{$\ \ \hat f_+^{{\cal I}^+}$};
\end{tikzpicture}~~~~~~~~~~~~~~~~~~~~~~~~~~~
\centering\begin{tikzpicture}[scale=.75]
\definecolor{darkgreen}{rgb}{.0, 0.5, .1};
\draw (0,6) 
-- (2.75,3.25 )node[above right ]{\!\! $\hat f_-^{{\cal I}^+}$} -- (3,3) -- (2.8,2.8) node[below right ]{\!\!\!\!\! $\hat f_+^{{\cal I}^-}$} --
(0,0)--
(0,0)node[right]{\, $\hat f^{\mathcal{I}^-}_-$} --  (0,6)node[right]{$\ \ \hat f_+^{{\cal I}^+}$};
\end{tikzpicture}
\vspace{-2mm}
\caption{Soft boundary modes associated with the null components of our in and out Cauchy slices for the gravitational collapse geometry (left) and for flat Minkowski space (right).  }
\label{fig:boundarydata}
\end{figure}

The soft boundary data for the two situations are indicated in Figure~\ref{fig:boundarydata}. Here we chose to represent the black hole space-time as a gravitational collapse geometry with a future horizon and no past horizon; and instead of using a transition function, we now use a single coordinate patch and represent the soft modes in terms of the shifted location of the black hole horizon. The boundary modes for past null infinity ${\cal I}^-$ and the future horizon ${\cal H}$  are identified with the respective supertranslation modes in the advanced Bondi gauge. The boundary modes at future null infinity ${\cal I}^+$ are defined as the supertranslation modes in the retarded Bondi gauge. Along all three null boundaries, the far future and far past values of the soft modes $f_\pm$ are functions of the transverse coordinates $z^A$ only and satisfy the commutation relations
\bea\label{chargecommutations}
\bigl[\hat f_+^{{\cal I}^-}(z), \hat{f}_-^{{\cal I}^-}(z')\bigr] \, = \, \bigl[\hat f_+^{{\cal I}^+}(z), \hat{f}_-^{{\cal I}^+}(z')\bigr]\, =\, \bigl[\hat f_+^{{\cal H}}(z), \hat{f}_-^{{\cal H}}(z')\bigr] \! \is   i\spc \Omega(z,z')
\eea
with other commutators vanishing. Here $\Omega(z,z')$ denotes the Green's function of the operator $D^2(D^2+2)$. Before imposing any constraint relations,  the soft modes of the collapsing geometry span a six-dimensional phase space for every transverse point $z$, while for Minkowski space the soft phase space at $z$ is four dimensional.

Charge conservation imposes a linear relation among the supertranslation generators acting on the corresponding regions of the form $
Q^{{\cal I}^-}_{f_1}\! =  Q^{{\cal H}}_{f_2} +Q^{{\cal I}^+}_{f_3}$, 
or written out as an integral over $z$
\bea
\label{chargec}
\int\! d^2 z \bigl(f_1(z)\, Q^{{\cal I}^-}\!\!\spc (z) - f_2(z) \, Q^{{\cal I}^+}\!\!\spc (z) - f_3(z) \, Q^{{\cal H}}\nspc (z) \bigr)\! \is\! 0.
\eea
Here the supertranslation parameters $f_1$, $f_2$ and $f_3$ are linearly related via propagation through the bulk. Within $\mathcal{S}$-matrix insertions, this is equation \eqref{qsplit}. 
The respective supertranslation currents split into a sum over a soft and a hard contribution as
\bea
Q^{{\cal I}^-} \!\!\! \is \textstyle\frac{1}{{ 16}\pi}  D^2 (D^2 +2)(\hat f_+^{{\cal I}^-}\!\nspc -{\hat f}^{{\cal I}^-}_-\nspc )\, +\, Q^{{\cal I}^-}_H\nonumber\\[1.5mm]
Q^{{\cal I}^+} \!\!\! \is   \textstyle\frac{1}{{ 16}\pi} D^2 (D^2 +2)(\spc {\hat f}_-^{{\cal I}^+}\!\nspc  -\spc \hat f_+^{{\cal I}^+} \nspc )\, + \, Q^{{\cal I}^+}_H\label{qshsplit}\\[1.5mm]
Q^{\cal H}\! \is \textstyle \frac{1}{{16}\pi}  D^2 (D^2 +2)(\spc {\hat f}^{\cal H}_-\spc -\spc {\hat f}^{\cal H}_+\spc )\; +\; Q_H^{\cal H}\nonumber .
\eea
Equation~\eqref{qshsplit} can be plugged into the charge conservation condition \eqref{chargec}.  It follows, from the commutation relations \eqref{chargecommutations} that 
the combination of charges in \eqref{chargec} generate a gauge symmetry under simultaneous shifts 
\bea
\label{fshift}
{\hat f}^{{\cal I}^-}_\pm \to {\hat f}^{{\cal I}^-}_\pm +  f_1, \qquad & \quad
{\hat f}^{{\cal I}^+}_\pm \to {\hat f}^{{\cal I}^+}_\pm +  f_2, \qquad & \quad
{\hat f}^{{\cal H}}_\pm \to {\hat f}^{{\cal H}}_\pm +  f_3. 
\eea 
Because~\eqref{chargec} is a first class constraint, it eliminates two degrees of freedom per $z$: the sum of memory modes appearing in \eqref{chargec} once \eqref{qshsplit} is plugged in, and the conjugate Goldstone mode involving a simultanous shift in the zero modes~\eqref{fshift}.

At the full non-linear level, the relation between supertranslation parameters $f_1$, $f_2$ and $f_3$ depends on the bulk matter flux encoded in the hard charges. At the linearized level, the propagation only depends on the classical background geometry. In either case, there is only one independent gauge function. Dividing out the symmetry, this leaves a four dimensional phase space at every transverse location in the case of Figure~\ref{fig:boundarydata}a, or a two dimension phase space in the case of no horizon Figure~\ref{fig:boundarydata}b.

In the pure Minkowski setting without a black hole there are no horizon modes.  In using advanced and retarded Bondi gauge for $\mathcal{I}^-$ and $\mathcal{I}^+$, respectively, the supertranslation charge conservation condition \eqref{chargec} simplifies to the special case with $f_3=0$ and $f_1=f_2$.  We can rearrange our soft sector phase space via 
\bea \begin{array}{cc}\label{independent}
\qquad \hat{f}^{\cal I^+}_{+}\!\!-\hat{f}^{\cal I^+}_{-}\!\! +\hat{f}^{\cal I^-}_{+}\!\! -\hat{f}^{\cal I^-}_{-},\ &\  \hat{f}^{\cal I^+}_{+}\!\! +\hat{f}^{\cal I^+}_{-}\!\! +\hat{f}^{\cal I^-}_{+}\!\! +\hat{f}^{\cal I^-}_{-}\,~~ \\[3.5mm]
\qquad \hat{f}^{\cal I^+}_{+}\!\!+\hat{f}^{\cal I^+}_{-}\!\! -\hat{f}^{\cal I^-}_{+}\!\! -\hat{f}^{\cal I^-}_{-},\ 
&\hat{f}^{\cal I^+}_{+}\!\!\!  - \hat{f}^{\cal I^+}_{-}\!\! -\hat{f}^{\cal I^-}_{+}\!\! +\hat{f}^{\cal I^-}_{-}.
\end{array}
\eea
The first and second rows are mutually commuting, while the first entry of the first row is the combination of soft modes appearing in~\eqref{chargec}.  This first row is eliminated by gauge fixing the Goldstone mode.  

We will now see that the second row is also eliminated.
CPT invariance motivates the identification between the late supertranslation mode on ${\cal I}^-$ and the early supertranslation mode on ${\cal I}^+$
\bea\label{cpt}
\hat f_-^{{\cal I}^+} \!\! \is \! \hat f_+^{{\cal I}^-} \qquad \quad \Leftrightarrow \quad \begin{array} {c} {\mbox{ CPT\ invariance\ condition\ at}}\\[1mm]
{\mbox{spatial\ infinity\ in\ Minkowski\  space .}}\end{array}
\eea
The difference between the entries in the first column of~\ref{independent} is proportional to $f_-^{{\cal I}^+} - \hat f_+^{{\cal I}^-}$. The soft phase space is thus eliminated in Minkowski space after imposing gauge invariance and the CPT relation.

A similar analysis can be performed in the presence of a horizon.  
The six dimensional phase space of Figure~\ref{fig:boundarydata}a gets reduced to four dimensions after gauge fixing.
In the case of the gravitational collapse geometry, there is not a good justification for imposing CPT invariance in this direct way.  However, if one argues that a matching condition near spatial infinity should not be insensitive to the interior, and proceeds to impose~\eqref{cpt}, one is left with a non-trivial two dimensional soft phase space.  In the presence of the black hole horizon, there are two  soft hair modes that are not determined by the hard charges. According to the HPS proposal, these soft modes store the soft part $S_{so\! f\!t}$ of the hidden entropy of the black hole. Semi-classically, we may equate $S_{so\! f \! t}$ with the phase space volume of the soft hair phase space, measured in Planck units.

\bibliographystyle{utphys}
\bibliography{references}

\end{document}